\documentclass[preprint,prd,aps,showpacs,showkeys,onecolumn,english]{revtex4}
\usepackage{amsmath,amssymb} 
\usepackage[dvips]{graphicx} 
\usepackage{graphics}
\usepackage{graphicx}
\usepackage{epsfig}
\usepackage{slashed}
\usepackage[english]{babel}
\usepackage{amsmath}
\usepackage{amssymb}
\usepackage{graphics}
\usepackage{graphicx}
\usepackage{epsfig}
\usepackage{slashed}

\newcommand\nn{\nonumber}
\newcommand\ba{\begin{eqnarray}}
\newcommand\ea{\end{eqnarray}}

\begin{document}
\title{Muon pair production in the process $e^+ +e^- \to e^+ +e^- +\mu^+ + \mu^-$ by the two-photon collisions $\gamma \gamma \to \mu^+ \mu^-$ at high energies.
The transverse momentum distributions}
\author{A. I.~Ahmadov$^{a,b}$ \footnote{E-mail: ahmadov@theor.jinr.ru}}
\affiliation{$^{a}$ Bogoliubov Laboratory of Theoretical Physics,
JINR, Dubna, 141980 Russia}
\affiliation{$^{b}$ Institute of Physics, Azerbaijan
National Academy of Sciences, H.Javid ave. 131, AZ-1143 Baku, Azerbaijan}

\date{\today}

\begin{abstract}

In this present paper, we investigate the muon pairs production in the interaction between two quasireal photons in $e^+e^-$ collision.
The total and differential cross section of the process $\gamma \gamma \to \mu^+\mu^-$ at a beam energy of photons from 3 GeV to 40 GeV in the center-of-mass and for different values of muon transverse momentum and the muon rapidity and the muon angle are calculated.
We also study the total cross section,  as a function of the $e^+ e^-$ center-of-mass energy $\sqrt {s}$ in the region
5 GeV $\leq \sqrt {s} \leq$ 209 GeV process of the $e^+ +e^- \to e^+ + e^- +\mu^+ + \mu^-$ by the two-photon mechanism.
The obtained our results are in satisfactory agreement with the experimental data.

\vspace*{0.5cm}

\noindent
\end{abstract}

\maketitle
\section{Introduction}
\label{Introduction}

In the microworld physics have a mathematically consistent theory -the standard model, one of the main problems is the confirmation of this model and the search for possible exits beyond its limits, and successfully tested in detail at colliders, accelerators \cite{Glashow,Weinberg,Salam1,Salam2}. \\
In particle physics are continually searching the most fundamental method for the description matter and the forces, which can be govern. \\
The main task in physics are describe all physical phenomena with a few basic principles.
Elementary particle physics by means of quarks and leptons, and their interactions tries to explain the variety of the Universe. \\
Quantum Electrodynamics (QED) and Standard Model (SM), had already been successful in remodeling the electromagnetic and weak interaction between
charged particles and of intermediate bosons, and photons. \\
The electron-positron, photon-photon and photon-electron scattering processes in QED and SM at colliders of high energies have large attention
both theoretical and experimental.
The exists accelerators with high-energy colliding $e^+e^-$, $\gamma\gamma$, $\gamma e$, $\mu^+\mu^-$ beams wide use to study the fundamental
interactions \cite{E1,E2,E3,E4,E5,E6,E7,E8,E9,E10,E11,E12,E13,E14}. The some QED of processes maybe play an important role at these colliders,
which the with increasing energy the cross section does not drop.

The laws of physics in the nuclear domain is for the most part derived from analyzing the outcomes of the particles collisions at high-energy. \\
For the study of photon-photon collisions are of the main physics goals of investigation at the high energy electron-positron linear collider. \\
Among the fundamental predictions of Standard Model are of the investigations with large transverse momentum and angle in exclusive processes.  \\
The massive lepton-pair production in photon-photon and hadron-hadron collisions are studied already for many years. \\
There is an experiments, that measured the production of a muon pair and is fairly well described in QED.
This to make it important to study the process of muon pair production in lepton-lepton collisions. \\
Photon-photon processes producing a large transverse momentum  muon pair play a fundamental role in electron-positron  collisions.
Lepton pair production in photon-photon collisions is intimately related to many signals and their backgrounds for new physics. \\
The necessary noted, that two-photon processes $\gamma\gamma \to X$ are investigation in detail and is an important section of modern high-energy physics
\cite{Budnev,Morgan,Isaev}.
In the experiment, they are studied on counter $e^+e^-$ beams at interacting virtual photons, which the are emitted by the initial particles
(that is, in the reaction $e^+e^- \to e^+e^-\gamma^* \gamma^* \to e^+e^- X$).

The study of two-photon collisions are arises interest in exclusive production of hadron pairs at high momentum transfer in the framework of
perturbative Quantum Chromodynamics \cite{P1,P2,P3,P4}.
Therefore, two photon interactions can provide an important information for some effects in Quantum Chromodynamics. \\
The reaction $e^+e^- \to e^+e^- \ell^+\ell^-$, where $\ell$ can be any charged lepton, which the arises from $O (\alpha^4)$  detailed investigate in \cite{Budnev}. \\
Thus, at $\alpha^4$ of the order the investigation of the process $e^+e^- \to e^+e^-\mu^+\mu^-$ provides a good way to test QED. \\
Namely, for the study of the processes $e^+e^- \to e^+e^-\mu^+\mu^-$ are well suited the LEP collider, which the have high precision detectors \cite{L31,L32}.  \\
Electron-positron colliders are ideal for study the characteristics of lepton pairs productions, because the production mechanisms involve the electromagnetic and electroweak interactions. Therefore, the cross-sections can be estimated quite reliably, as can final-state distributions.

Up to now many studies have been carried out for processes photon-photon collisions both theoretical and experimental,  which the have been made important
contributions in this areas \cite{Budnev,Morgan,gamma1,gamma11,gamma12,gamma13,gamma14,gamma15,gamma16,gamma17,gamma18,gamma19,
gamma20,gamma21,gamma22,gamma23,gamma24,gamma25,gamma26,gamma27,gamma28,gamma29,gamma30,gamma31,gamma32,gamma33,gamma34,gamma35,gamma36,gamma37,gamma38,
gamma39,gamma40,gamma41,gamma42,gamma43,gamma44,gamma45,gamma46,gamma47,gamma48,gamma49,gamma50,gamma51,gamma52,gamma53,gamma54,gamma55,gam1,gam2,gam3,gam4,
gam5,gam6,gam7,gam8,gam9,gam10,gam11,gam12,gam13,gam14,gam15,gam16,gam17,gam18,gam19,gam20,gam21,gam22,gam23,gam24,gam25,gam26,gam27,gam28,gam29,gam30,gam31,
gam32,gam33,gam34,gam35,gam36,gam37,gam38,gamexp1,gamexp2,gamexp3,gamexp4,gamexp5,gamexp6,gamexp7,gamexp8,gamexp9,gamexp10,gamexp11,gamexp12,gamexp13,L3,OPAL}. \\
For the study of the process $e^+ +e^- \to e^+ +e^- +\mu^+ + \mu^-$ provides was carried out several experiment \cite{L3,OPAL,eeexp1,eeexp2,eeexp3,eeexp4,eeexp5,eeexp6},
and are have made theoretical calculations within Quantum Electrodynamics.
There are many groups have made relevant calculations \cite{Budnev,eeeemm1,eeeemm2,eeeemm3,eeeemm4,eeeemm5,eeeemm6,eeeemm7,eeeemm8,eeeemm9,
eeeemm10,eeeemm11,eeeemm12,eeeemm13,eeeemm14,eeeemm15,eeeemm16,eeeemm17,eeeemm18,eeeemm19,eeeemm20,eeeemm21,eeeemm22,eeeemm23}. \\
In the main, a large number of studies of the photon-photon collisions processes have been performed at the detectors at LEP, at BELLE, at BABAR, at VEPP, and
other $e^+e^-$ colliders. In the L3 detector at LEP have been studied the process $e^+ +e^- \to e^+ +e^- +\mu^+ + \mu^-$ at $161 \,\,GeV < \sqrt {s} <209\,\,GeV$.
In this process the muon pair invariant mass for the photon-photon collisions was measured in the range $3\,\,GeV < W_{\gamma\gamma} < 40\,\,GeV$ \cite{L3}.

In this paper, we shall calculate total and differential cross sections for $\gamma\gamma \to \mu^+ \mu^-$ (and $e^+ + e^- \to e^+ e^- \mu^+ \mu^-$)
from $\sqrt {s}$, $p_T$, rapidity ($y$) and angle within the framework of QED (and SM). \\
Also, we will be study of the $e^+ +e^- \to e^+ + e^- +\mu^+ + \mu^-$ process, which the initiated by electron-positron collisions, and that,
illustrate some fundamental concepts and issues central to the QED and SM. \\
We want to note that the center-of-mass energies are in the range of the LEP collider $161 \,\,GeV \leq \sqrt {s} \leq \,\,209 \,\,GeV $, and the scattering angle are in the range $15^{\circ}< \theta < 165^{\circ}$.

\section{The process $\bf {\gamma\gamma \to \mu^+ \mu^-}$}
\label{GF}

\begin{figure}[!htb]
\includegraphics[width=0.43\linewidth]{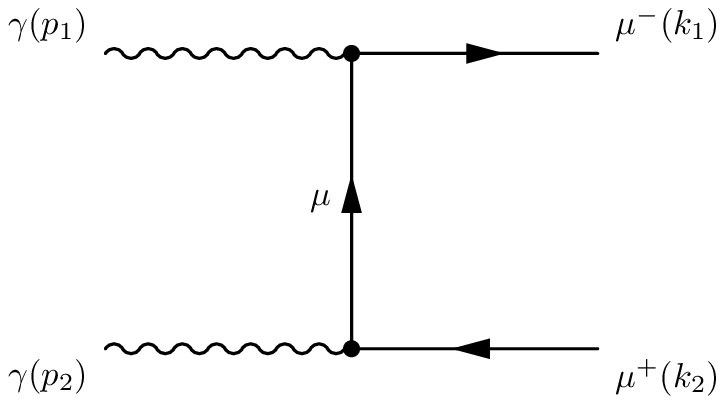}
\hspace*{1cm}
\includegraphics[width=0.43\linewidth]{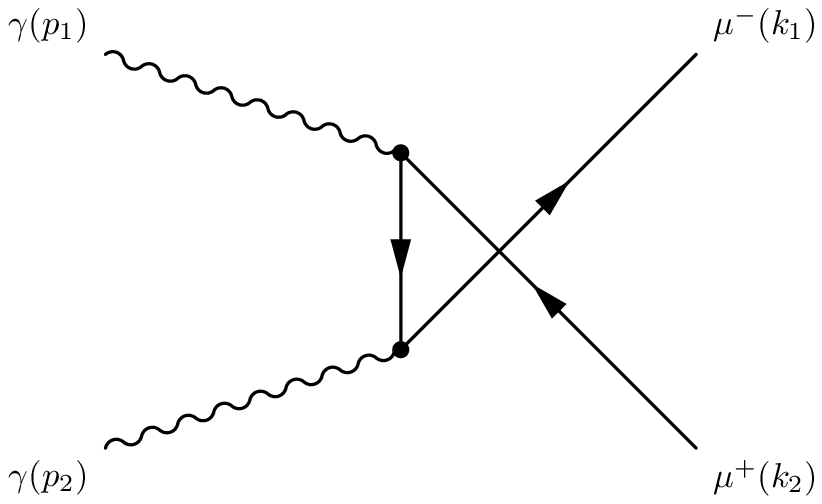}
\parbox{20mm} { (1) }
\parbox{30mm} {~~~ }
\parbox{50mm} { \qquad \qquad  \qquad  (2) }
       \caption{The Feynman diagrams for the muon pairs production in $\gamma\gamma$ collisions in the Born approximation.}
       \label{diagram1}
\end{figure}

In the present paper, in electron-positron collisions, massive muon-pair production proceeds through photon-photon collisions,
i.e. through two-photon mechanism. \\
To study the $e^+ +e^- \to e^+ + e^- +\mu^+ + \mu^-$ process, at the first stage we must consider the production of muon pairs
in $\gamma \gamma \to \mu^+ \mu^-$ collisions.
Therefore, in this section we want to discuss the $\mu^+\mu^-$ production in gamma-gamma collisions. \\
This process can be represented in the following physical picture: at leading order, two photons interact and produce a measured final state,
of mass $m_{\mu}$ and transverse momentum $p_T$. \\
The process is written in the form
\ba
\gamma(p_1) + \gamma(p_2) \to \mu^-(k_1) + \mu^+(k_2).
\label{gamma}
\ea
The Feynman diagrams of the process \eqref{gamma} in the Born approximation is shown in Fig.~\ref{diagram1}
%

\begin{figure}[!htb]
       \includegraphics[width=0.43\linewidth]{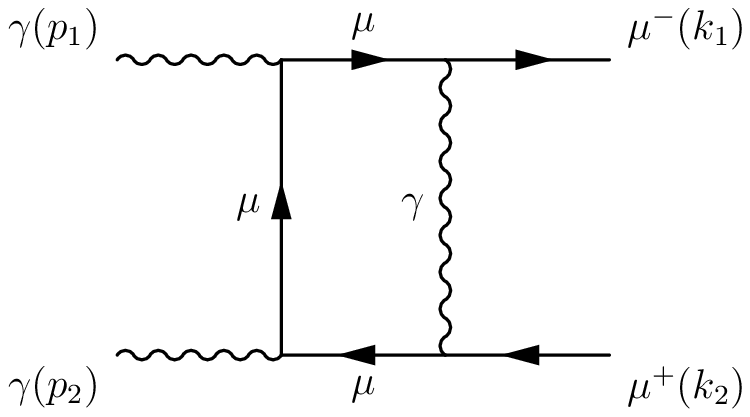}
\hspace*{1cm}
       \includegraphics[width=0.43\linewidth]{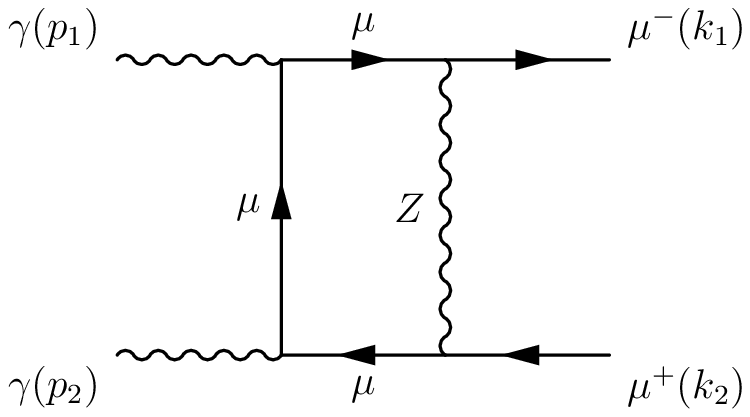}
\parbox{20mm} { (3) }
\parbox{30mm} {~~~ }
\parbox{50mm} { \qquad \qquad  \qquad  (4) }

\bigskip
\bigskip

       \includegraphics[width=0.43\linewidth]{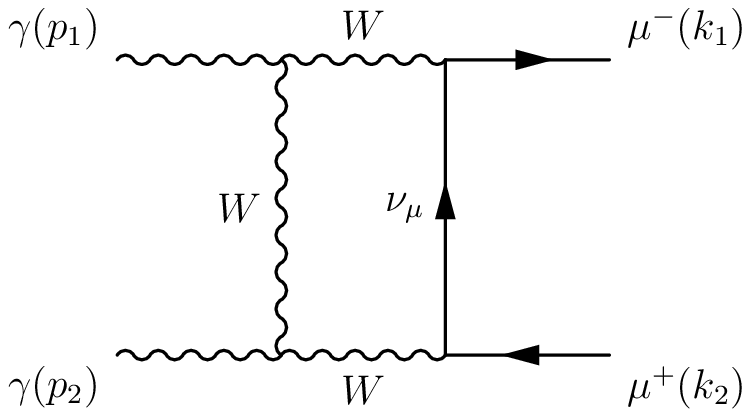}
\hspace*{1cm}
       \includegraphics[width=0.43\linewidth]{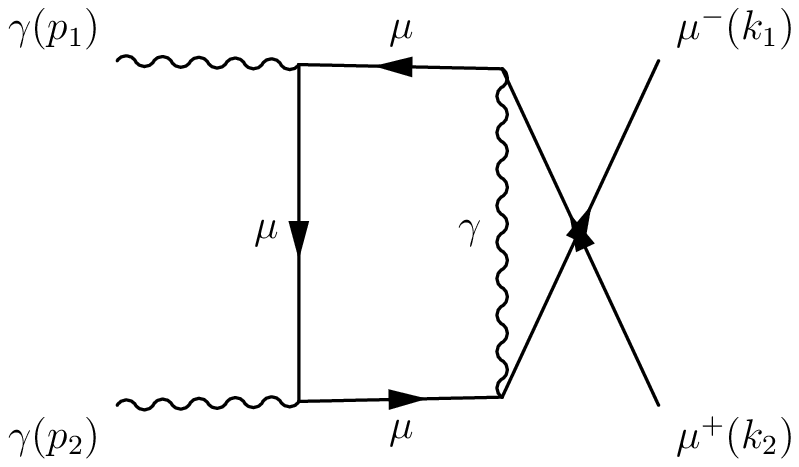}
\parbox{20mm} { (5) }
\parbox{30mm} {~~~ }
\parbox{50mm} { \qquad \qquad  \qquad  (6) }

\bigskip
\bigskip
      \includegraphics[width=0.43\linewidth]{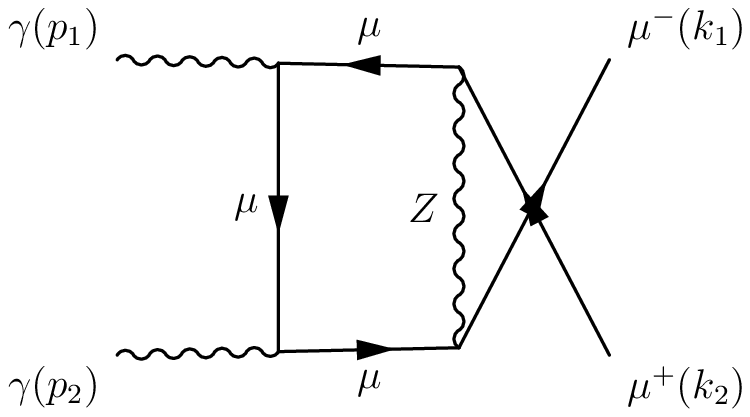}
\hspace*{1cm}
       \includegraphics[width=0.43\linewidth]{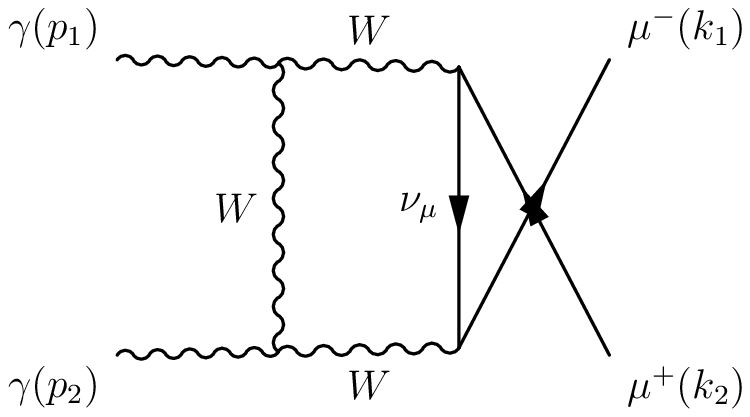}
\parbox{20mm} { (7) }
\parbox{30mm} {~~~ }
\parbox{50mm} { \qquad \qquad  \qquad  (8) }
       \caption{Box-type of the Feynman diagrams which contribute to the muon pair production in the process $\gamma\gamma \to \mu^+ \mu^-$ at one-loop level.}
       \label{diagram2}
\end{figure}

The kinematics of the process  \eqref{gamma} can be described in terms of the  following Mandelstam variables:
\ba
s = (p_1 + p_2)^2 = (k_1 + k_2)^2;  \nn \\
t = (p_1 - k_1)^2 = (p_2 - k_2)^2;  \nn \\
u = (p_1 - k_2)^2 = (p_2 - k_1)^2.
\label{Mand1}
\ea
In this process, the created muons have masses $m_{\mu}$, then it follows from \eqref{Mand1} that the sum of the Mandelstam variables, which we will use can be written in this form
\ba
s + t + u = 2m_{\mu}^2.
\label{Man2}
\ea
The matrix elements corresponding for the Feynman diagrams in the Born approximation (Fig.~\ref{diagram1})  we can write the form
%
\ba
{\it i}{\mathcal M}_1 &=& -\bar{u}(k_1, m_{\mu})(-ie\gamma_{\alpha})\frac{i(\hat{p}_2 - \hat{k}_2 +m_{\mu})}{(k_2 - p_2)^2 -m_{\mu}^2}(-ie\gamma_{\beta})v(k_2,m_{\mu})\varepsilon_{\alpha}(p_1)
\varepsilon_{\beta}(p_2);  \nn \\
{\it i}{\mathcal M}_2 &=& -\bar{u}(k_1, m_{\mu})(-ie\gamma_{\beta})\frac{i(\hat{k}_1 - \hat{p}_2 +m_{\mu})}{(p_2 - k_1)^2 -m_{\mu}^2}(-ie\gamma_{\alpha})v(k_2,m_{\mu})\varepsilon_{\alpha}(p_1)
\varepsilon_{\beta}(p_2);
\ea
The matrix elements for the Box diagrams (Fig.~\ref{diagram2}) we can write as follows:
\ba
{\it i}{\mathcal M}_3 &=&-\frac{i (2ie)^4}{16\pi^4}\int d^4q_1 \frac{\bar {u}(k_1)\gamma_{\mu}(\hat {k}_1+\hat {k}_2-\hat {p}_2-\hat {q}_1+m_{\mu})\gamma_{\alpha}
(-\hat {q}_1+m_{\mu})}{(q_1^2-m_{\mu}^2)((p_2+q_1-k_1-k_2)^2-m_{\mu}^2)} \cdot \nn \\
&& \cdot \frac{\gamma_{\beta}(-\hat {p}_2-\hat{q}_1+m_{\mu})\gamma_{\nu} v(k_2)}
{((p_2+q_1)^2-m_{\mu}^2)(p_2+q_1-k_2)^2}\cdot g^{\mu\nu}\varepsilon_{\alpha}(p_1)\,\,\varepsilon_{\beta}(p_2); \nn \\
{\it i}{\mathcal M}_4 &=&-\frac{i (2ie)^2}{16\pi^4}\int d^4q_1 \frac{\bar {u}(k_1)\gamma_{\mu}(\hat {k}_1+\hat {k}_2-\hat {p}_2-\hat {q}_1+m_{\mu})\gamma_{\alpha}
(-\hat {q}_1+m_{\mu})}{(q_1^2-m_{\mu}^2)((p_2+q_1-k_1-k_2)^2-m_{\mu}^2)} \cdot \nn \\
&& \cdot \frac{\gamma_{\beta}(-\hat {p}_2-\hat{q}_1+m_{\mu})\gamma_{\nu}v(k_2) \cdot g^{\mu\nu} \cdot \varepsilon_{\alpha}(p_1)\,\,\varepsilon_{\beta}(p_2)}
{((p_2+q_1)^2-m_{\mu}^2)((p_2+q_1-k_2)^2-M_Z^2)}\cdot \biggl(\frac{ie (-\frac{1}{2}+sin^2\theta_{W})}{cos\theta_W \,\,sin\theta_W} +\frac{ie sin\theta_W}{cos\theta_W}\biggr)^2;
\nn \\
{\it i}{\mathcal M}_5 &=&-\frac{ie^2}{16\pi^4}\int d^4q_1 \frac{\bar {u}(k_1)\gamma_{\sigma}(\hat {p}_2+\hat {q}_1-\hat {k}_2)\gamma_{\kappa}\cdot v(k_2) \cdot g^{\mu\nu} g^{\rho\sigma}g^{\lambda\kappa}} {(q_1^2-M_W^2)((p_2+q_1)^2-M_W^2)(p_2+q_1+k_2)^2((p_2+q_1-k_1-k_2)^2-M_W^2)}\cdot   \nn \\
&&\cdot \biggl(\frac{ie}{\sqrt {2}sin\theta_W}\biggr)^2 [(-p_1-q_1)^{\rho}
g^{\alpha\mu}+(p_1-p_2-q_1+k_1+k_2)^{\mu}g^{\alpha\rho}+(p_2+2q_1-k_1-k_2)^{\alpha}g^{\mu\rho}]\cdot \nn \\
&& \cdot [(-p_2+q_1)^{\lambda}g^{\beta\nu}+(2p_2+q_1)^{\nu}g^{\beta\lambda}+
(-p_2-2q_1)^{\beta}g^{\nu\lambda}]\cdot \varepsilon_{\alpha}(p_1)\,\,\varepsilon_{\beta}(p_2); \nn  \\
{\it i}{\mathcal M}_6 &=&-\frac{i(2ie)^4}{16\pi^4}\int d^4q_1 \frac{\bar {u}(k_1)\gamma_{\mu}(\hat {p}_2+\hat {q}_1+m_{\mu})\gamma_{\beta}
(\hat {q}_1+m_{\mu})\gamma_{\alpha}}{(q_1^2-m_{\mu}^2)((p_2+q_1-k_1-k_2)^2-m_{\mu}^2)} \cdot \nn \\
&& \cdot \frac{(\hat {p}_2+\hat{q}_1 -\hat {k}_1-\hat {k}_2+m_{\mu})\gamma_{\nu}v(k_2)}
{((p_2+q_1)^2-m_{\mu}^2)(p_2+q_1-k_2)^2}\cdot g^{\mu\nu}\varepsilon_{\alpha}(p_1)\,\,\varepsilon_{\beta}(p_2); \nn
\ea
\ba
{\it i}{\mathcal M}_7 &=&-\frac{i (2ie)^2}{16\pi^4}\int d^4q_1 \frac{\bar {u}(k_1)\gamma_{\mu}(\hat {p}_2 +\hat{q}_1+m_{\mu})\gamma_{\beta}
(\hat {q}_1+m_{\mu})\gamma_{\alpha}(\hat {p}_2 +\hat {q}_1 -\hat {k}_1 -\hat {k}_2+m_{\mu})}{(q_1^2-m_{\mu}^2)((p_2+q_1-k_1-k_2)^2-m_{\mu}^2)((p_2+q_1)^2-m_{\mu}^2)} \cdot \nn \\
&& \cdot \frac{\gamma_{\nu}v(k_2) \cdot g^{\mu\nu} \cdot \varepsilon_{\alpha}(p_1)\,\,\varepsilon_{\beta}(p_2)}
{((p_2+q_1-k_2)^2-M_Z^2)}\cdot \biggl(\frac{ie (-\frac{1}{2}+sin^2\theta_{W})}{cos\theta_W \,\,sin\theta_W} +\frac{ie sin\theta_W}{cos\theta_W}\biggr)^2;
\nn \\
{\it i}{\mathcal M}_8 &=&-\frac{ie^2}{16\pi^4}\int d^4q_1 \frac{\bar {u}(k_1)\gamma_{\lambda}(-\hat {p}_2-\hat {q}_1+\hat {k}_1)\gamma_{\sigma}\cdot v(k_2) \cdot g^{\mu\nu} g^{\rho\sigma}g^{\lambda\kappa}}{(q_1^2-M_W^2)((p_2+q_1)^2-M_W^2)(p_2+q_1-k_1)^2((p_2+q_1-k_1-k_2)^2-M_W^2)}\cdot   \nn \\
&&\cdot \biggl(\frac{ie}{\sqrt {2}sin\theta_W}\biggr)^2 [(-p_1-q_1)^{\rho}
g^{\alpha\mu}+(p_1-p_2-q_1+k_1+k_2)^{\mu}g^{\alpha\rho}+(p_2+2q_1-k_1-k_2)^{\alpha}g^{\mu\rho}]\cdot \nn \\
&& \cdot [(-p_2+q_1)^{\kappa}g^{\beta\nu}+(2p_2+q_1)^{\nu}g^{\beta\kappa}+
(-p_2-2q_1)^{\beta}g^{\nu\kappa}]\cdot \varepsilon_{\alpha}(p_1)\,\,\varepsilon_{\beta}(p_2);
\ea
We will be consider to the subprocess $\gamma\gamma \to \mu^+ \mu^-$ and, in the leading order, and also in the second order of perturbation theory.
The total matrix element of the process $\gamma\gamma \to \mu^+ \mu^-$ will has the form
\ba
M_{\gamma\gamma \to \mu^+ \mu^-} = M_{Born} + M_{\text{Box}},
\label{TM}
\ea
where the $M_{Born}$ is the leading order, the $M_{\text{Box}}$ is the second order amplitude. \\
Then, the general matrix elements of the Feynman diagrams in the Born approximation in Fig.~\ref{diagram1} can be written in the following form
\ba
M_{Born} = {\mathcal M}_1 + {\mathcal M}_2.
\label{TM1}
\ea
The general matrix elements for the Box diagrams in Fig.~\ref{diagram2} can be written as follows
\ba
M_{\text{Box}} = {\mathcal M}_3 + {\mathcal M}_4 + {\mathcal M}_5 + {\mathcal M}_6 + {\mathcal M}_7 + {\mathcal M}_8.
\label{TM2}
\ea
We will not take into account of the higher order $\alpha^2$ in the Box diagram.
Then the square of the full amplitude summed over the polarizations of particles, can be written in the following form
\ba
M = \frac{1}{4} \sum_{pol}|M_{\gamma\gamma \to \mu^+ \mu^-}|^2 \approx \frac{1}{4} \sum_{pol}|M_{\text{Born}}|^2  + \frac{1}{2} \sum_{pol} Re(M_{Born}^+ \cdot M_{\text{Box}}).
\ea

In case taking into account the mass of muons, $k_1^2 = k_2^2=m^2_{\mu}$  (for the process $\gamma\gamma \to \mu^+ \mu^-$), we can obtain the border
equation for the physical regions in the following form
\ba
t u = m_{\mu}^4.
\label{tu}
\ea
For a givens $s$ and mass muon  after solving two equations \eqref{Man2} and \eqref{tu}, we can define the upper and lower bounds for $t$, is defined as
\ba
t^{\text {max, min}} = \frac{-s+2m_{\mu}^2}{2} \pm \frac{s}{2}\sqrt{1-\frac{4m_{\mu}^2}{s}}.
\label{tpm}
\ea
After integration in \eqref{TM1} by $t$, according to formula \eqref{tpm}, for the square of the matrix element of the Born diagram, we obtain the following expression:
\ba
&&\overline{|M_{Born}|^2} = (4\pi\alpha)^2 \biggl\{(2 s m_{\mu}^2 - 10 m_{\mu}^4) \cdot \frac{\sqrt{s(s-4m_{\mu}^2)}}{s m_{\mu}^2} - 2 s \cdot \biggl(\frac{\sqrt{s(s-4m_{\mu}^2)}}{s} + \nn \\
&& +\ln\frac{s-\sqrt{s(s-4m_{\mu}^2)}}{s+\sqrt{s(s-4m_{\mu}^2)}}\biggr) -
\frac{2}{s}(s+m_{\mu}^2)\sqrt{s(s-4m_{\mu}^2)} + \biggl(6 s m_{\mu}^2 - 18 m_{\mu}^4 \biggr)\cdot \frac{\sqrt{s(s-4m_{\mu}^2)}}{s m_{\mu}^2} + \nn \\
&&+(8 m_{\mu}^2 -2s)\cdot \biggl(\frac{m_{\mu}^2 -s}{m_{\mu}^2 s}\sqrt{s(s-4m_{\mu}^2)} +\ln\frac{s+\sqrt{s(s-4m_{\mu}^2)}}{s-\sqrt{s(s-4m_{\mu}^2)}}\biggr)- \nn \\
&&- 2 \biggl(\frac{1}{s m_{\mu}^2}(m_{\mu}^4 - s m_{\mu}^2 +s^2)\sqrt{s(s-4m_{\mu}^2)} + 2(m_{\mu}^2 -s) \ln\frac{s+\sqrt{s(s-4m_{\mu}^2)}}{s-\sqrt{s(s-4m_{\mu}^2)}}\biggr) + \nn \\
&&+ \frac{1}{s} \biggl(8 s m_{\mu}^2 - 12 m_{\mu}^4 \biggr)\biggl(\ln\frac{s+\sqrt{s(s-4m_{\mu}^2)}}{s-\sqrt{s(s-4m_{\mu}^2)}} + \ln\frac{-s-\sqrt{s(s-4m_{\mu}^2)}}{-s+\sqrt{s(s-4m_{\mu}^2)}}\biggr)\biggr\}.
\ea
The terms $M_{Born}^+ \cdot M_{\text{Box}}$ can be written in the following form:
\ba
&&{\mathcal M}_1^+ \cdot {\mathcal M}_3 = -\frac{(4\pi \alpha)^3}{\pi^4} \frac{g_{\mu\nu} g_{\alpha\alpha'} g_{\beta\beta'}}{(k_2-p_2)^2-m_{\mu}^2}
\int d^4q_1 \frac{Tr (\hat {k}_2 -m_{\mu})\gamma_{\beta'}(\hat {p}_2 -\hat {k}_2 +m_{\mu})\gamma_{\alpha'}(\hat {k}_1 +m_{\mu})\gamma_{\mu}}
{(q_1^2 -m_{\mu}^2)((p_2 +q_1)^2-m_{\mu}^2)} \cdot \nn \\
&&\cdot \frac{(-\hat {p}_2 -\hat{q}_1 +\hat {k}_1 +\hat {k}_2 +m_{\mu})\gamma_{\alpha} (-\hat {q}_1 +m_{\mu})\gamma_{\beta} (-\hat {p}_2 -\hat {q}_1 +m_{\mu})\gamma_{\nu}}
{(p_2 +q_1 -k_2)^2 ((p_2 +q_1 -k_1 -k_2)^2 -m_{\mu}^2)}, \nn \\
&&{\mathcal M}_1^+ \cdot {\mathcal M}_4 = -\frac{16 G_F^3 sin^6\theta_W M^6_W}{\pi^4} \biggl(\frac{sin^2\theta_W -\frac{1}{2}}{cos\theta_W sin\theta_W} +\frac{sin\theta_W}{cos\theta_W}\biggr)^2 \frac{g_{\mu\nu} g_{\alpha\alpha'} g_{\beta\beta'}}{(k_2-p_2)^2-m_{\mu}^2} \cdot \nn \\
&&\cdot \int d^4q_1 \frac{Tr (\hat {k}_2 -m_{\mu})\gamma_{\beta'}(\hat {p}_2 -\hat {k}_2 +m_{\mu})\gamma_{\alpha'}(\hat {k}_1 +m_{\mu})\gamma_{\mu}}
{(q_1^2 -m_{\mu}^2)((p_2 +q_1)^2-m_{\mu}^2) ((p_2 +q_1 -k_2)^2 -M^2_Z)} \cdot \nn \\
&&\cdot \frac{(-\hat {p}_2 -\hat{q}_1 +\hat {k}_1 +\hat {k}_2 +m_{\mu})\gamma_{\alpha} (-\hat {q}_1 +m_{\mu})\gamma_{\beta} (-\hat {p}_2 -\hat {q}_1 +m_{\mu})\gamma_{\nu}}
{((p_2 +q_1 -k_1 -k_2)^2 -m_{\mu}^2)}, \nn \\
&&{\mathcal M}_1^+ \cdot {\mathcal M}_5 = \frac{(4\pi \alpha)^2 M^2_W G_F}{4 \sqrt {2} \pi^4} \frac{g_{\mu\nu} g_{\rho\sigma} g_{\lambda\kappa} g_{\alpha\alpha'} g_{\beta\beta'}}{(k_2-p_2)^2-m_{\mu}^2} \cdot
\int d^4q_1 \frac{Tr (\hat {k}_2 -m_{\mu})\gamma_{\beta'}(\hat {p}_2 -\hat {k}_2 +m_{\mu})\gamma_{\alpha'}}
{(q_1^2 -M_W^2)((p_2 +q_1)^2-M_W^2)} \cdot \nn \\
&&\cdot \frac{(\hat {k}_1 +m_{\mu})\gamma_{\lambda}(-\hat {p}_2 -\hat{q}_1 +\hat {k}_1)\gamma_{\sigma}}{((p_2 +q_1 -k_1 -k_2)^2 -M_W^2)(p_2 +q_1 -k_2)^2} \cdot
\biggl((-p_1-q_1)^{\rho}g_{\alpha\mu} +(p_1 -p_2 -q_1 +k_1 +k_2)^{\mu}g_{\alpha\rho} + \nn \\
&&+(p_2+2q_1-k_1-k_2)^{\alpha}g_{\mu\rho}\biggr)\cdot
\biggl((-p_2+q_1)^{\kappa}g_{\beta\nu} + (2p_2+q_1)^{\nu}g_{\beta\kappa} +(-p_2-2q_1)^{\beta}g_{\nu\kappa}\biggr),  \nn  \\
&&{\mathcal M}_1^+ \cdot {\mathcal M}_6 = \frac{(4\pi \alpha)^3}{\pi^4} \frac{g_{\mu\nu} g_{\alpha\alpha'} g_{\beta\beta'}}{(k_2-p_2)^2-m_{\mu}^2}
\int d^4q_1 \frac{Tr (\hat {k}_2 -m_{\mu})\gamma_{\beta'}(\hat {p}_2 -\hat {k}_2 +m_{\mu})\gamma_{\alpha'}(\hat {k}_1 +m_{\mu})\gamma_{\mu}}
{(q_1^2 -m_{\mu}^2)((p_2 +q_1)^2-m_{\mu}^2)} \cdot \nn \\
&&\cdot \frac{(\hat {p}_2 +\hat{q}_1 +m_{\mu})\gamma_{\beta} (\hat {q}_1 +m_{\mu})\gamma_{\alpha} (\hat {p}_2 +\hat {q}_1-\hat {k}_1 -\hat {k}_2 +m_{\mu})\gamma_{\nu}}
{(p_2 +q_1 -k_1)^2 ((p_2 +q_1 -k_1 -k_2)^2 -m_{\mu}^2)}, \nn
\ea
\ba
&&{\mathcal M}_1^+ \cdot {\mathcal M}_7 = -\frac{16 G_F^3 sin^6\theta_W M^6_W}{\pi^4} \biggl(\frac{sin^2\theta_W -\frac{1}{2}}{cos\theta_W sin\theta_W} +\frac{sin\theta_W}{cos\theta_W}\biggr)^2 \frac{g_{\mu\nu} g_{\alpha\alpha'} g_{\beta\beta'}}{(k_2-p_2)^2-m_{\mu}^2} \cdot \nn \\
&&\cdot \int d^4q_1 \frac{Tr (\hat {k}_2 -m_{\mu})\gamma_{\beta'}(\hat {p}_2 -\hat {k}_2 +m_{\mu})\gamma_{\alpha'}(\hat {k}_1 +m_{\mu})\gamma_{\mu}}
{(q_1^2 -m_{\mu}^2)((p_2 +q_1)^2-m_{\mu}^2) ((p_2 +q_1 -k_1)^2 -M^2_Z)} \cdot \nn \\
&&\cdot \frac{(\hat {p}_2 +\hat{q}_1 +m_{\mu})\gamma_{\beta} (\hat {q}_1 +m_{\mu})\gamma_{\alpha} (\hat {p}_2 +\hat {q}_1-\hat {k}_1-\hat {k}_2 +m_{\mu})\gamma_{\nu}}
{((p_2 +q_1 -k_1 -k_2)^2 -m_{\mu}^2)}, \nn \\
&&{\mathcal M}_1^+ \cdot {\mathcal M}_8 = \frac{(4\pi \alpha)^2 M^2_W G_F}{4 \sqrt {2} \pi^4} \frac{g_{\mu\nu} g_{\rho\sigma} g_{\lambda\kappa} g_{\alpha\alpha'} g_{\beta\beta'}}{(k_2-p_2)^2-m_{\mu}^2} \cdot
\int d^4q_1 \frac{Tr (\hat {k}_2 -m_{\mu})\gamma_{\beta'}(\hat {p}_2 -\hat {k}_2 +m_{\mu})\gamma_{\alpha'}}
{(q_1^2 -M_W^2)((p_2 +q_1)^2-M_W^2)} \cdot \nn \\
&&\cdot \frac{(\hat {k}_1 +m_{\mu})\gamma_{\lambda}(-\hat {p}_2 -\hat{q}_1 +\hat {k}_1)\gamma_{\sigma}}{((p_2 +q_1 -k_1 -k_2)^2 -M_W^2)(p_2 +q_1 -k_1)^2} \cdot
\biggl((-p_1-q_1)^{\rho}g_{\alpha\mu} +(p_1 -p_2 -q_1 +k_1 +k_2)^{\mu}g_{\alpha\rho} + \nn \\
&&+(p_2+2q_1-k_1-k_2)^{\alpha}g_{\mu\rho}\biggr)\cdot
\biggl((-p_2+q_1)^{\kappa}g_{\beta\nu} + (2p_2+q_1)^{\nu}g_{\beta\kappa} +(-p_2-2q_1)^{\beta}g_{\nu\kappa}\biggr),  \nn  \\
&&{\mathcal M}_2^+ \cdot {\mathcal M}_3 = -\frac{(4\pi \alpha)^3}{\pi^4} \frac{g_{\mu\nu} g_{\alpha\alpha'} g_{\beta\beta'}}{(p_2-k_1)^2-m_{\mu}^2}
\int d^4q_1 \frac{Tr (\hat {k}_2 -m_{\mu})\gamma_{\alpha'}(\hat {k}_1 -\hat {p}_2 +m_{\mu})\gamma_{\beta'}(\hat {k}_1 +m_{\mu})\gamma_{\mu}}
{(q_1^2 -m_{\mu}^2)((p_2 +q_1)^2-m_{\mu}^2)} \cdot \nn \\
&&\cdot \frac{(-\hat {p}_2 -\hat{q}_1 +\hat {k}_1 +\hat {k}_2 +m_{\mu})\gamma_{\alpha} (-\hat {q}_1 +m_{\mu})\gamma_{\beta} (-\hat {p}_2 -\hat {q}_1 +m_{\mu})\gamma_{\nu}}
{(p_2 +q_1 -k_2)^2 ((p_2 +q_1 -k_1 -k_2)^2 -m_{\mu}^2)}, \nn \\
&&{\mathcal M}_2^+ \cdot {\mathcal M}_4 = -\frac{16 G_F^3 sin^6\theta_W M^6_W}{\pi^4} \biggl(\frac{sin^2\theta_W -\frac{1}{2}}{cos\theta_W sin\theta_W} +\frac{sin\theta_W}{cos\theta_W}\biggr)^2 \frac{g_{\mu\nu} g_{\alpha\alpha'} g_{\beta\beta'}}{(p_2-k_1)^2-m_{\mu}^2} \cdot \nn \\
&&\cdot \int d^4q_1 \frac{Tr (\hat {k}_2 -m_{\mu})\gamma_{\alpha'}(\hat {k}_1 -\hat {p}_2 +m_{\mu})\gamma_{\beta'}(\hat {k}_1 +m_{\mu})\gamma_{\mu}}
{(q_1^2 -m_{\mu}^2)((p_2 +q_1)^2-m_{\mu}^2) ((p_2 +q_1 -k_2)^2 -M^2_Z)} \cdot \nn \\
&&\cdot \frac{(-\hat {p}_2 -\hat{q}_1 +\hat {k}_1 +\hat {k}_2 +m_{\mu})\gamma_{\alpha} (-\hat {q}_1 +m_{\mu})\gamma_{\beta} (-\hat {p}_2 -\hat {q}_1 +m_{\mu})\gamma_{\nu}}
{((p_2 +q_1 -k_1 -k_2)^2 -m_{\mu}^2)}, \nn \\
&&{\mathcal M}_2^+ \cdot {\mathcal M}_5 = \frac{(4\pi \alpha)^2 M^2_W G_F}{4 \sqrt {2} \pi^4} \frac{g_{\mu\nu} g_{\rho\sigma} g_{\lambda\kappa} g_{\alpha\alpha'} g_{\beta\beta'}}{(p_2-k_1)^2-m_{\mu}^2} \cdot
\int d^4q_1 \frac{Tr (\hat {k}_2 -m_{\mu})\gamma_{\alpha'}(\hat {k}_1 -\hat {p}_2 +m_{\mu})\gamma_{\beta'}}
{(q_1^2 -M_W^2)((p_2 +q_1)^2-M_W^2)} \cdot \nn \\
&&\cdot \frac{(\hat {k}_1 +m_{\mu})\gamma_{\sigma}(\hat {p}_2 +\hat{q}_1 -\hat {k}_2)\gamma_{\kappa}}{((p_2 +q_1 -k_1 -k_2)^2 -M_W^2)(p_2 +q_1 -k_2)^2} \cdot
\biggl((-p_1-q_1)^{\rho}g_{\alpha\mu} +(p_1 -p_2 -q_1 +k_1 +k_2)^{\mu}g_{\alpha\rho} + \nn \\
&&+(p_2+2q_1-k_1-k_2)^{\alpha}g_{\mu\rho}\biggr)\cdot
\biggl((-p_2+q_1)^{\lambda}g_{\beta\nu} + (2p_2+q_1)^{\nu}g_{\beta\lambda} +(-p_2-2q_1)^{\beta}g_{\nu\lambda}\biggr),  \nn  \\
&&{\mathcal M}_2^+ \cdot {\mathcal M}_6 = \frac{(4\pi \alpha)^3}{\pi^4} \frac{g_{\mu\nu} g_{\alpha\alpha'} g_{\beta\beta'}}{(p_2-k_1)^2-m_{\mu}^2}
\int d^4q_1 \frac{Tr (\hat {k}_2 -m_{\mu})\gamma_{\alpha'}(\hat {k}_1 -\hat {p}_2 +m_{\mu})\gamma_{\beta'}(\hat {k}_1 +m_{\mu})\gamma_{\mu}}
{(q_1^2 -m_{\mu}^2)((p_2 +q_1)^2-m_{\mu}^2)} \cdot \nn \\
&&\cdot \frac{(\hat {p}_2 +\hat{q}_1 +m_{\mu})\gamma_{\beta} (\hat {q}_1 +m_{\mu})\gamma_{\alpha} (\hat {p}_2 +\hat {q}_1-\hat {k}_1 -\hat {k}_2 +m_{\mu})\gamma_{\nu}}
{(p_2 +q_1 -k_1)^2 ((p_2 +q_1 -k_1 -k_2)^2 -m_{\mu}^2)}, \nn \\
&&{\mathcal M}_2^+ \cdot {\mathcal M}_7 = -\frac{16 G_F^3 sin^6\theta_W M^6_W}{\pi^4} \biggl(\frac{sin^2\theta_W -\frac{1}{2}}{cos\theta_W sin\theta_W} +\frac{sin\theta_W}{cos\theta_W}\biggr)^2 \frac{g_{\mu\nu} g_{\alpha\alpha'} g_{\beta\beta'}}{(p_2-k_1)^2-m_{\mu}^2} \cdot \nn \\
&&\cdot \int d^4q_1 \frac{Tr (\hat {k}_2 -m_{\mu})\gamma_{\alpha'}(\hat {k}_1 -\hat {p}_2 +m_{\mu})\gamma_{\beta'}(\hat {k}_1 +m_{\mu})\gamma_{\mu}}
{(q_1^2 -m_{\mu}^2)((p_2 +q_1)^2-m_{\mu}^2) ((p_2 +q_1 -k_1)^2 -M^2_Z)} \cdot \nn \\
&&\cdot \frac{(\hat {p}_2 +\hat{q}_1 +m_{\mu})\gamma_{\beta} (\hat {q}_1 +m_{\mu})\gamma_{\alpha} (\hat {p}_2 +\hat {q}_1-\hat {k}_1-\hat {k}_2 +m_{\mu})\gamma_{\nu}}
{((p_2 +q_1 -k_1 -k_2)^2 -m_{\mu}^2)}, \nn
\ea
\ba
&&{\mathcal M}_2^+ \cdot {\mathcal M}_8 = \frac{(4\pi \alpha)^2 M^2_W G_F}{4 \sqrt {2} \pi^4} \frac{g_{\mu\nu} g_{\rho\sigma} g_{\lambda\kappa} g_{\alpha\alpha'} g_{\beta\beta'}}{(p_2-k_1)^2-m_{\mu}^2} \cdot
\int d^4q_1 \frac{Tr (\hat {k}_2 -m_{\mu})\gamma_{\alpha'}(\hat {k}_1 -\hat {p}_2 +m_{\mu})\gamma_{\beta'}}
{(q_1^2 -M_W^2)((p_2 +q_1)^2-M_W^2)} \cdot \nn \\
&&\cdot \frac{(\hat {k}_1 +m_{\mu})\gamma_{\lambda}(-\hat {p}_2 -\hat{q}_1 +\hat {k}_1)\gamma_{\sigma}}{((p_2 +q_1 -k_1 -k_2)^2 -M_W^2)(p_2 +q_1 -k_1)^2} \cdot
\biggl((-p_1-q_1)^{\rho}g_{\alpha\mu} +(p_1 -p_2 -q_1 +k_1 +k_2)^{\mu}g_{\alpha\rho} + \nn \\
&&+(p_2+2q_1-k_1-k_2)^{\alpha}g_{\mu\rho}\biggr)\cdot
\biggl((-p_2+q_1)^{\kappa}g_{\beta\nu} + (2p_2+q_1)^{\nu}g_{\beta\kappa} +(-p_2-2q_1)^{\beta}g_{\nu\kappa}\biggr),
\ea
where the quantities $M_W$, $M_Z$, $m_{\mu}$, $e = \sqrt {4\pi\alpha}$, $G_F$, and $\theta_W$ are the W-boson mass, the Z-boson mass, the muon mass, the elementary electric charge ($\alpha \sim 1/137$),
the Fermi coupling constant, and the weak mixing angle, respectively.  \\
In our calculation, we have used the completeness relation of the sum over the photon polarizations in the initial state,
replacing by the expression
\ba
\sum_{\lambda}\varepsilon_{\mu}(p_1,\lambda)\varepsilon_{\nu}^{\ast}(p_1,\lambda) = -g_{\mu\nu},
\ea
where $\varepsilon (p_1,\lambda)$ and $\varepsilon (p_2,\lambda)$ are the incoming photon polarization vectors. \\
The Box diagrams of the process \eqref{gamma} is shown in Fig.~\ref{diagram2}.

\begin{figure}[!htb]
       \centering
       \includegraphics[width=0.8\linewidth]{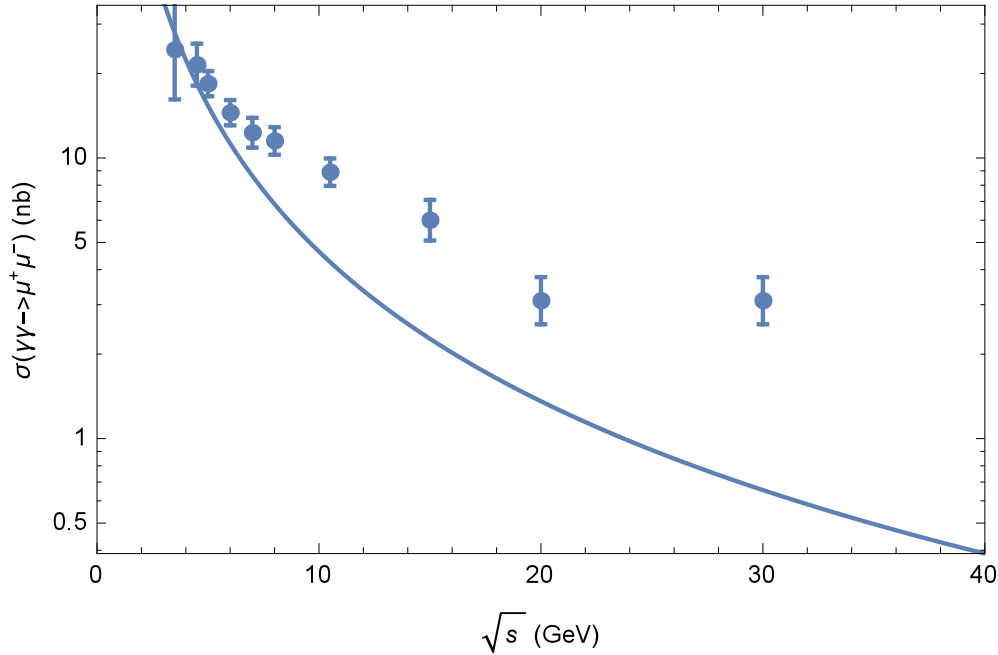}
       \caption{The total cross section of the process $\gamma \gamma \to \mu^+ \mu^-$ as a function of the $\gamma \gamma$ centre-of-mass energy.
       The experimental data is from \cite{L3}.}
       \label{sigma1}
\end{figure}

The "master formula" for evaluating the cross section and kinematic distributions for a $\gamma(p_1) + \gamma(p_2) \to \mu^-(k_1) + \mu^+(k_2)$ process is
\ba
d\sigma = \frac{1}{2s} \int \frac{d^3 \vec {k}_1}{(2\pi)^3 2E_1}\frac{d^3 \vec {k}_2}{(2\pi)^3 2E_2} (2\pi)^4 \delta^4(k_1 +k_2 - p_1 -p_2)\overline{|M|^2},
\label{DCS}
\ea
where $M$ is the invariant matrix element. \\
In phase volume after integrate over $\vec {k}_2$ using the $\delta$-function and  after some simple transformation, we can get differential
cross section for $\gamma\gamma \to \mu^+ \mu^-$ process \eqref{gamma}
in the center-of-mass frame in the following form:
\ba
\frac{d\sigma}{dt} = \frac{1}{16 \pi s^2} \beta_{\mu}\overline{|M|^2},
\label{ds}
\ea
where $\beta_{\mu} = \sqrt {1-\frac{4 m_{\mu}^2}{s}}$ and $\sqrt {s}$ is the total centre-of-mass energy of the colliding photons. \\
For obtaining the total cross section for the subprocess in photon-photon collisions \eqref{gamma} can be calculated by
\ba
\sigma (\gamma\gamma \to \mu^+ \mu^-) = \int\limits_{t^{\text {min}}}^{t^{\text {max}}} dt \frac{d\sigma}{dt},
\label{TCS}
\ea
where $t^{\text {max, min}}$ are defined in \eqref{tpm}.  \\
We are presenting the results of our numerical calculations in Fig.3 - Fig.7 and in Fig.9. \\
In accordance with experiment of L3 \cite{L3} we calculated total  cross section of the $\gamma\gamma \to \mu^+ \mu^-$
process \eqref{gamma} at center-of-mass energy in range $3\,\, GeV \leq \sqrt {s} \leq 40\,\, GeV$ and is plotted the dependence
of the total cross section from $\sqrt {s}$. We plot this dependence in Fig.~\ref{sigma1}.
It is shown that the cross section is slowly decreasing,  and   a  comparison with experimental data shows that our result
with a few experimental points are in good agreement, but full agreement we do not have.

\vspace*{0.05cm}
\section{The transverse momentum, the rapidity and angular distributions}

In this section we present the transverse-momentum, the rapidity and angular distributions for muon-pair production in photon-photon collisions.

It is well-known that particles (e.g. leptons), that produced in the final state should possess some intrinsic (primordial) transverse momentum.
It should be noted that, the recoil transverse momenta of dilepton pairs produced in hadronic collisions have been uniquely calculated in QCD
\cite{Politzer,Radyushkin,Fritzsch,Altarelli,Kajante1,Kajante2,Collins,Ravindran}.
Similarly, the recoil transverse momenta of muon pairs produced in photons collisions can be uniquely calculated in QED.  \\
It should also be noted, that intrinsic (primordial) transverse momenta of particles in initial state are almost negligible, therefore,
the experimentally observed large transverse momenta of massive muons produced in photons collisions have to arise almost entirely from the dynamical recoil effects. \\
Necessary be noted, that the transverse momentum spectrum of muon pairs production at lower mass $M_{\mu\mu}$ allows one to study in more detail
the non-perturbative contribution.
The measurements low-mass the process of the $\gamma\gamma \to \mu^+ \mu^-$ can be used to constrain the intrinsic transverse momentum distribution.
Therefore, process $\gamma\gamma \to \mu^+ \mu^-$, is one particularly interesting process, that produced massive lepton pairs of large transverse-momentum,
that can help to provide information on Standard Model. \\
It should be noted, that a precise measurement of the $p_T$ spectrum also provides an important input to the background prediction in searches of the
processes for beyond the Standard Model.

\begin{figure}[!htb]
       \centering
       \includegraphics[width=0.47\linewidth]{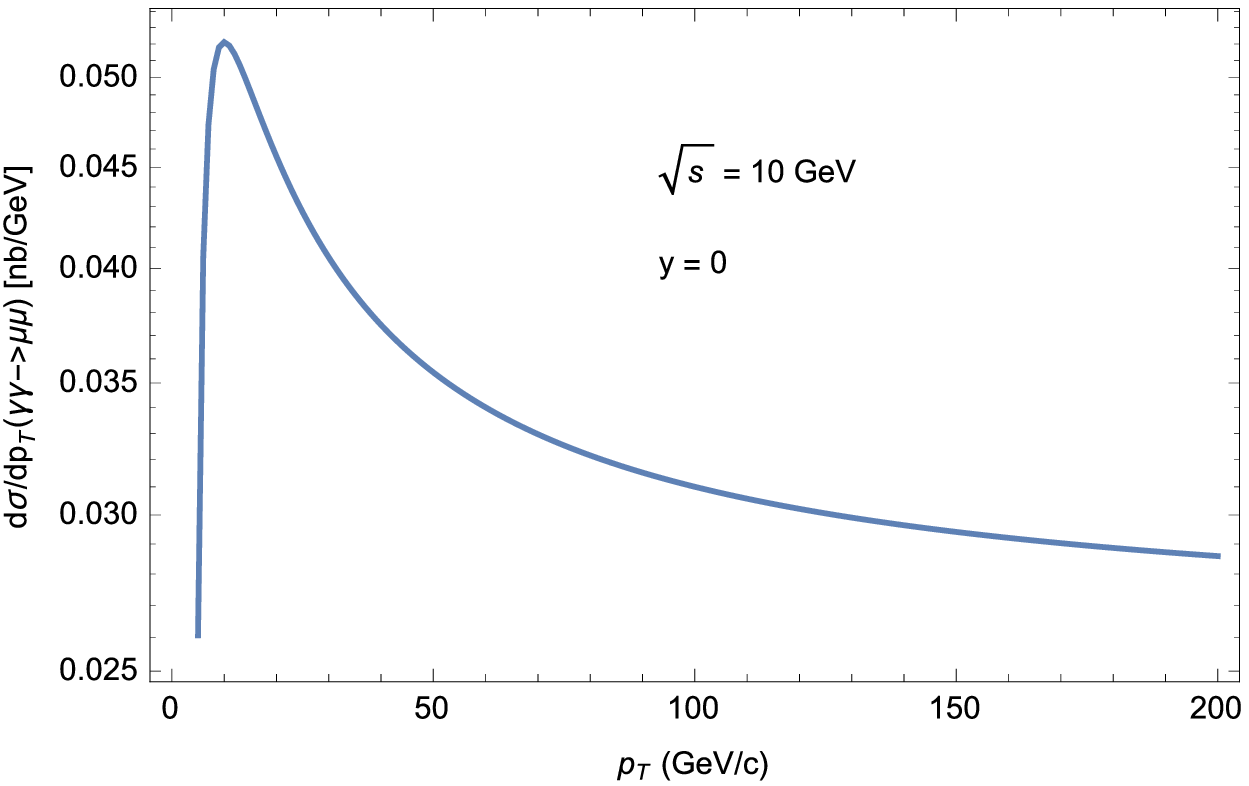}
       \includegraphics[width=0.47\linewidth]{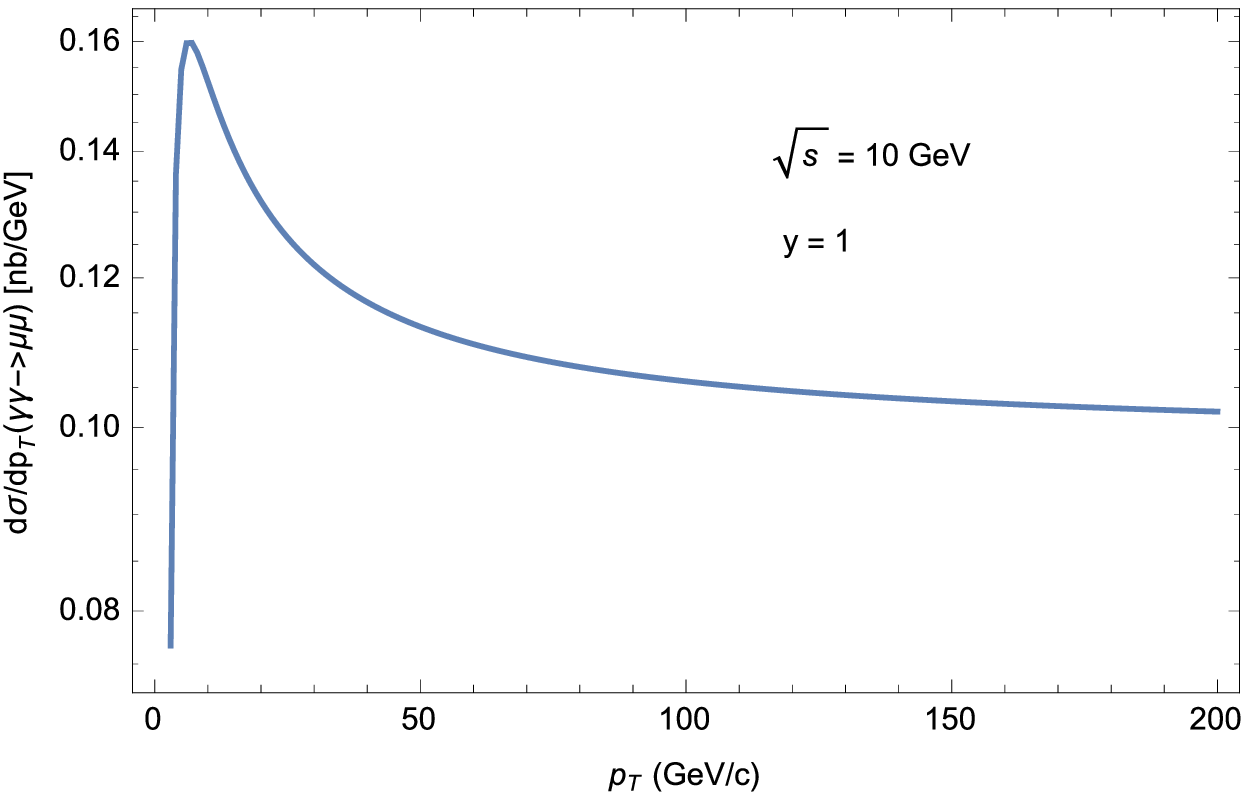} \\
\parbox{20mm} {\qquad \qquad (a) }
\parbox{9mm} {~~~}
\parbox{40mm} {\hspace*{0.15cm}} {\qquad \qquad (b) }

\vspace*{0.5cm}
    \hspace*{0.15cm}   \includegraphics[width=0.47\linewidth]{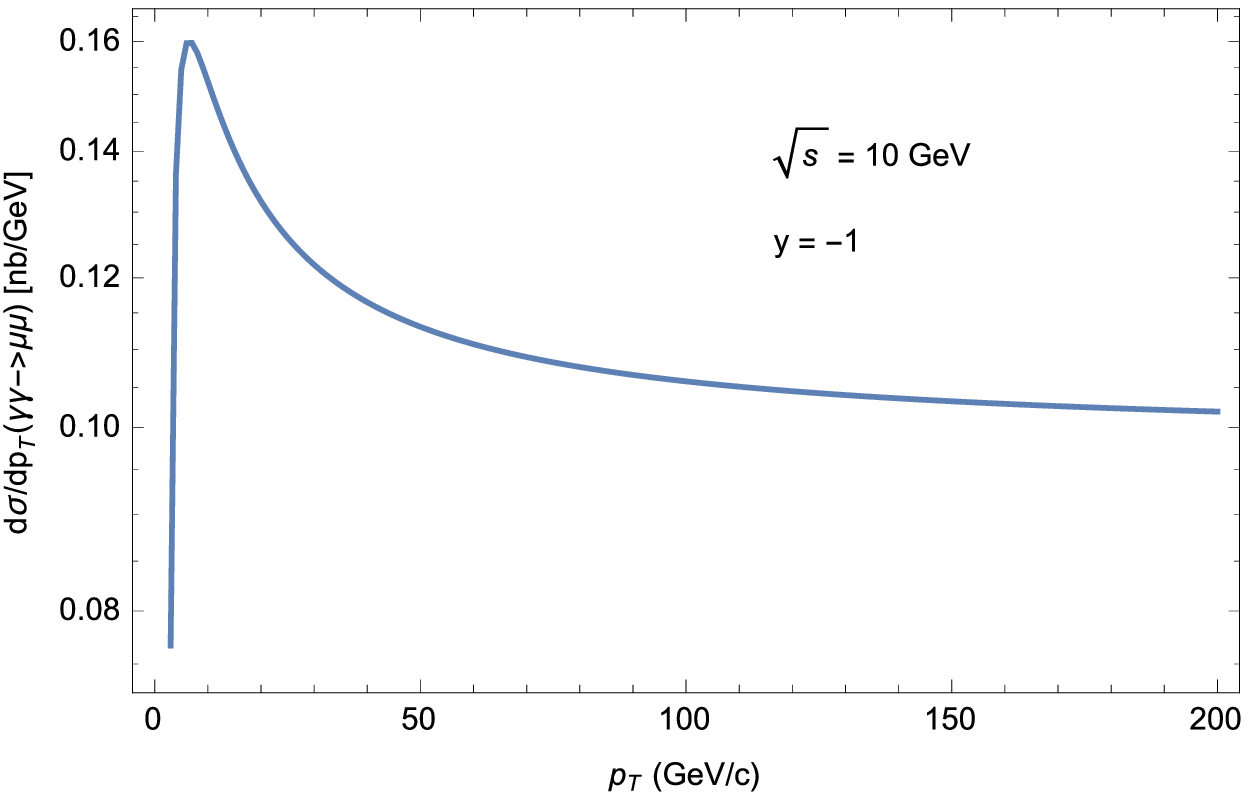}
     \hspace*{0.05cm}     \includegraphics[width=0.47\linewidth]{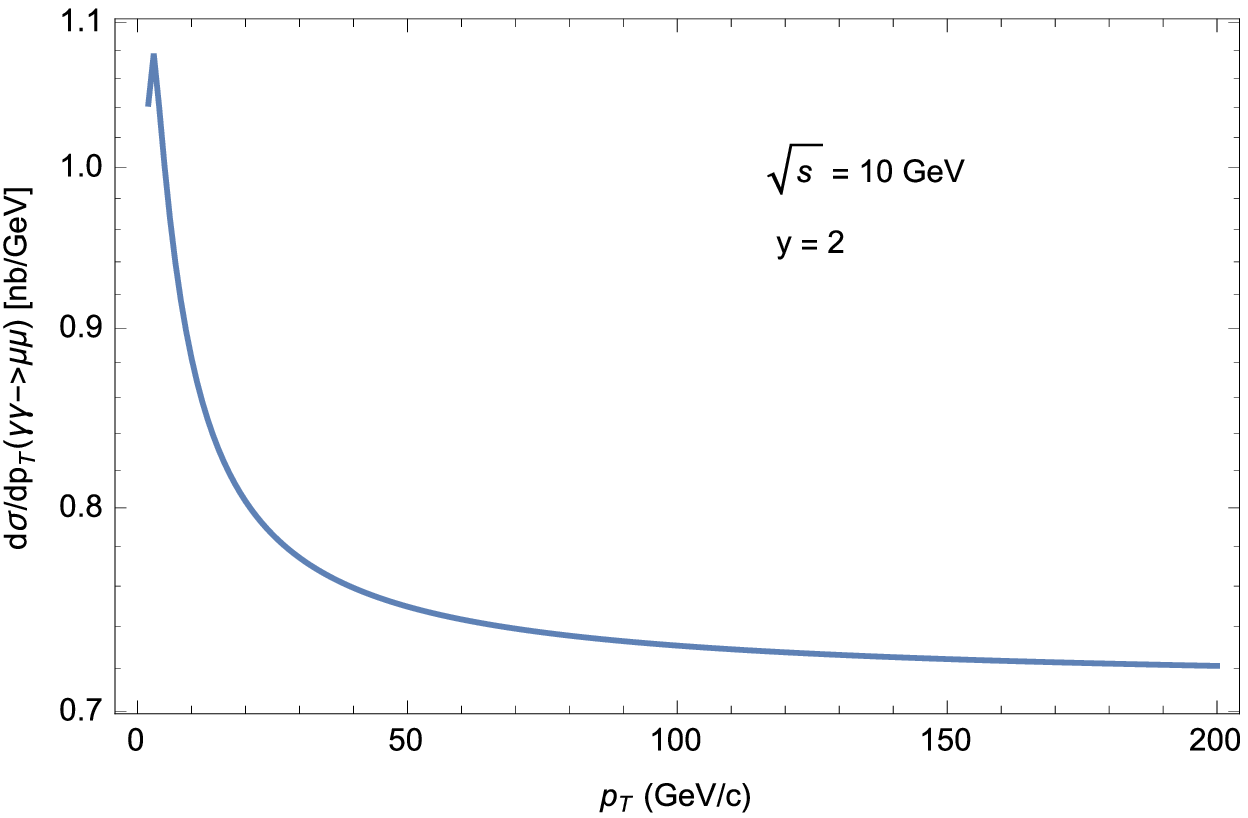}
\parbox{20mm} {\qquad \qquad (c) }
\parbox{9mm} {~~~}
\parbox{40mm} {\hspace*{0.15cm}} {\qquad \qquad (d) }
      \caption{Transverse momentum distribution of $\mu^{\pm}$ for fixed value of centre-of-mass energy $\sqrt {s}$ and for different values of $y$ rapidity. ($a$): $\sqrt {s}$ = 10 GeV, $y$ =0; \,\,($b$): $\sqrt {s}$ = 10 GeV, $y$ =1; \,\, ($c$): $\sqrt {s}$ = 10 GeV, $y$ = - 1;
      \,\, ($d$): $\sqrt {s}$ = 10 GeV, $y$ =2.}
       \label{pt1}
\end{figure}

\begin{figure}[!htb]
       \centering
       \includegraphics[width=0.47\linewidth]{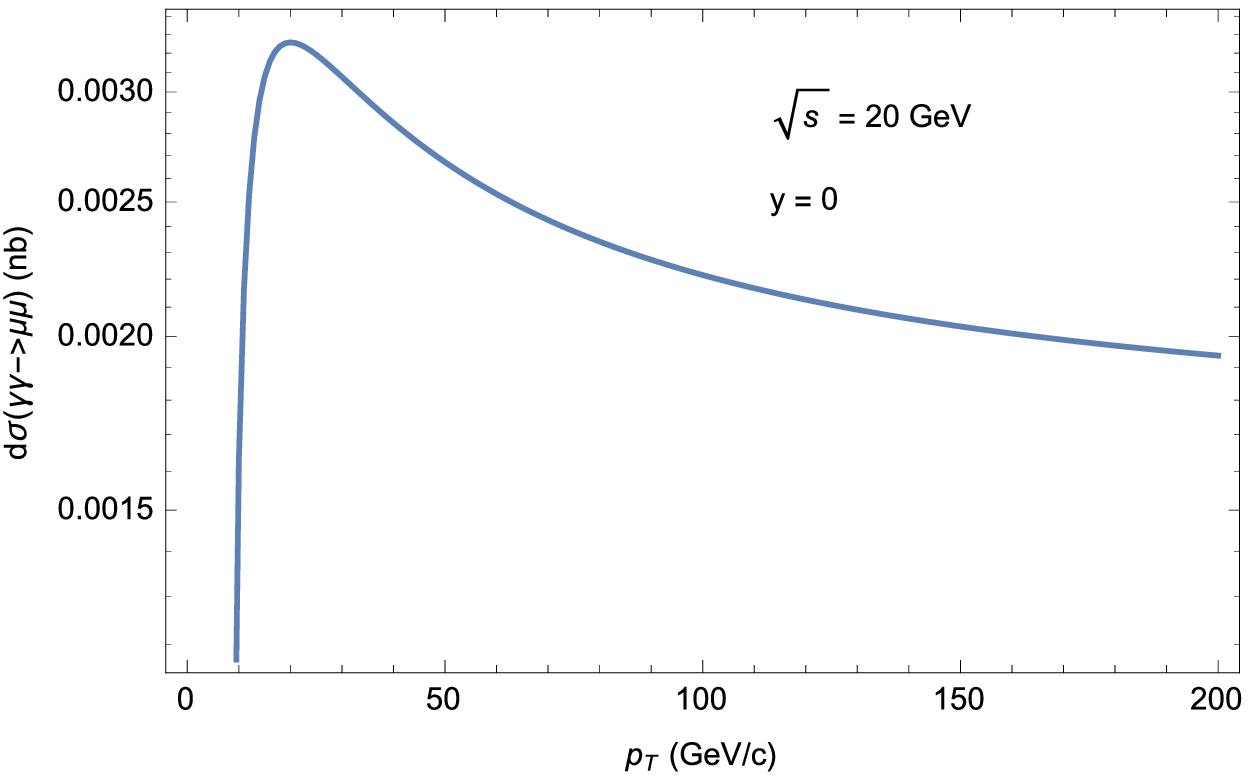}
       \includegraphics[width=0.47\linewidth]{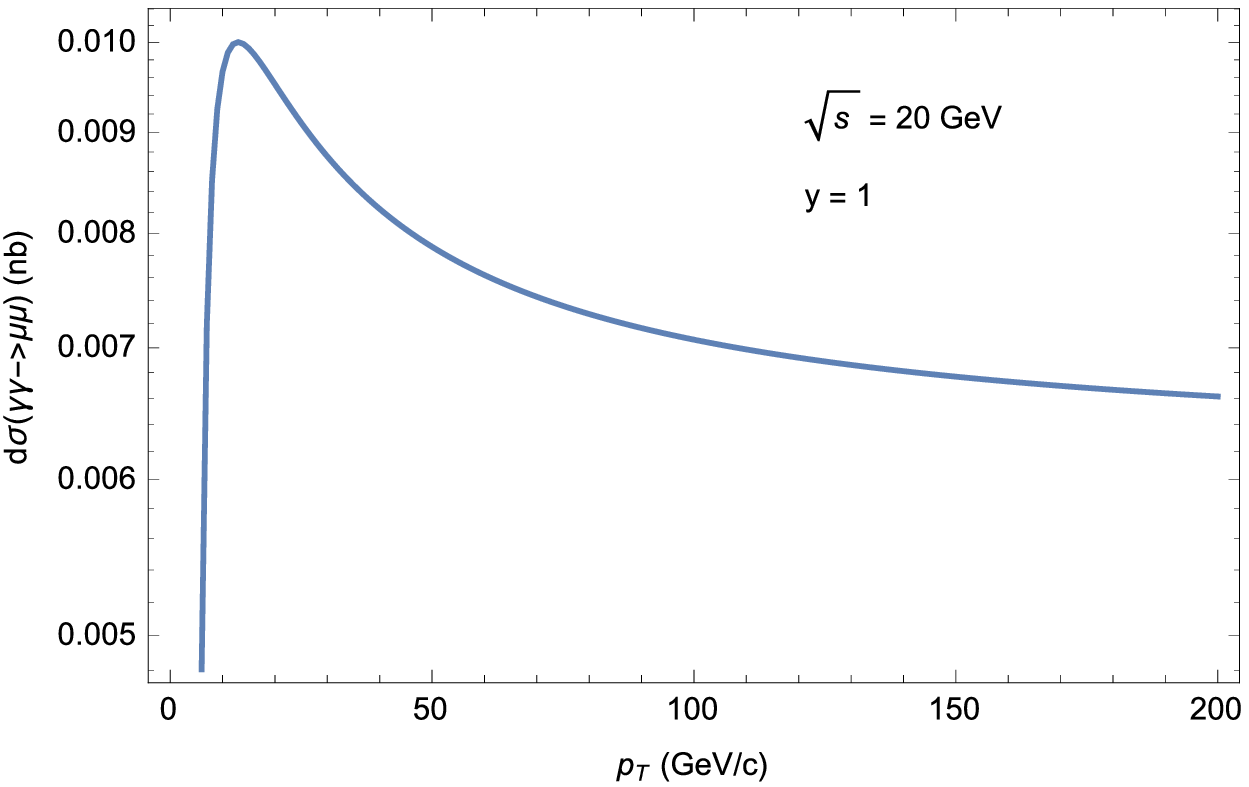} \\
\parbox{20mm} {\qquad \qquad (a) }
\parbox{9mm} {~~~}
\parbox{40mm} {\hspace*{0.15cm}} {\qquad \qquad (b) }

\vspace*{0.5cm}
    \hspace*{0.15cm}   \includegraphics[width=0.47\linewidth]{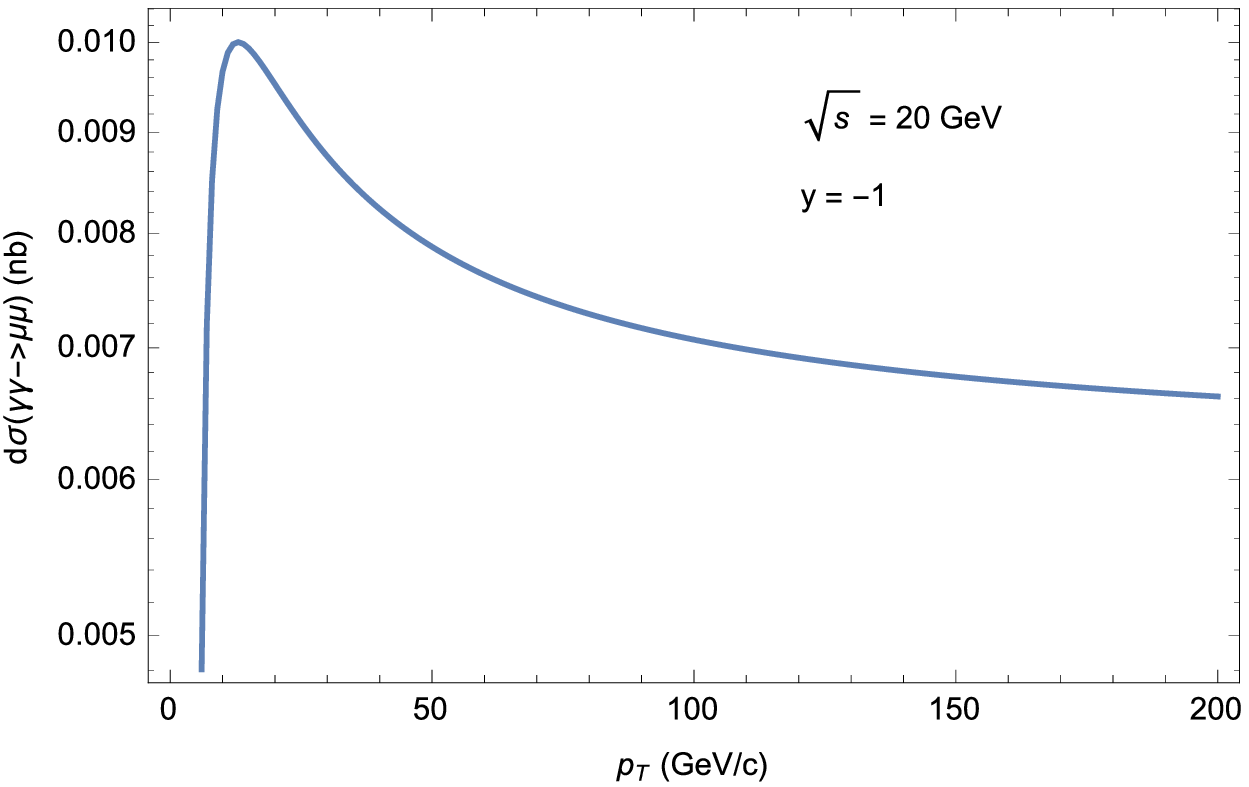}
     \hspace*{0.05cm}     \includegraphics[width=0.47\linewidth]{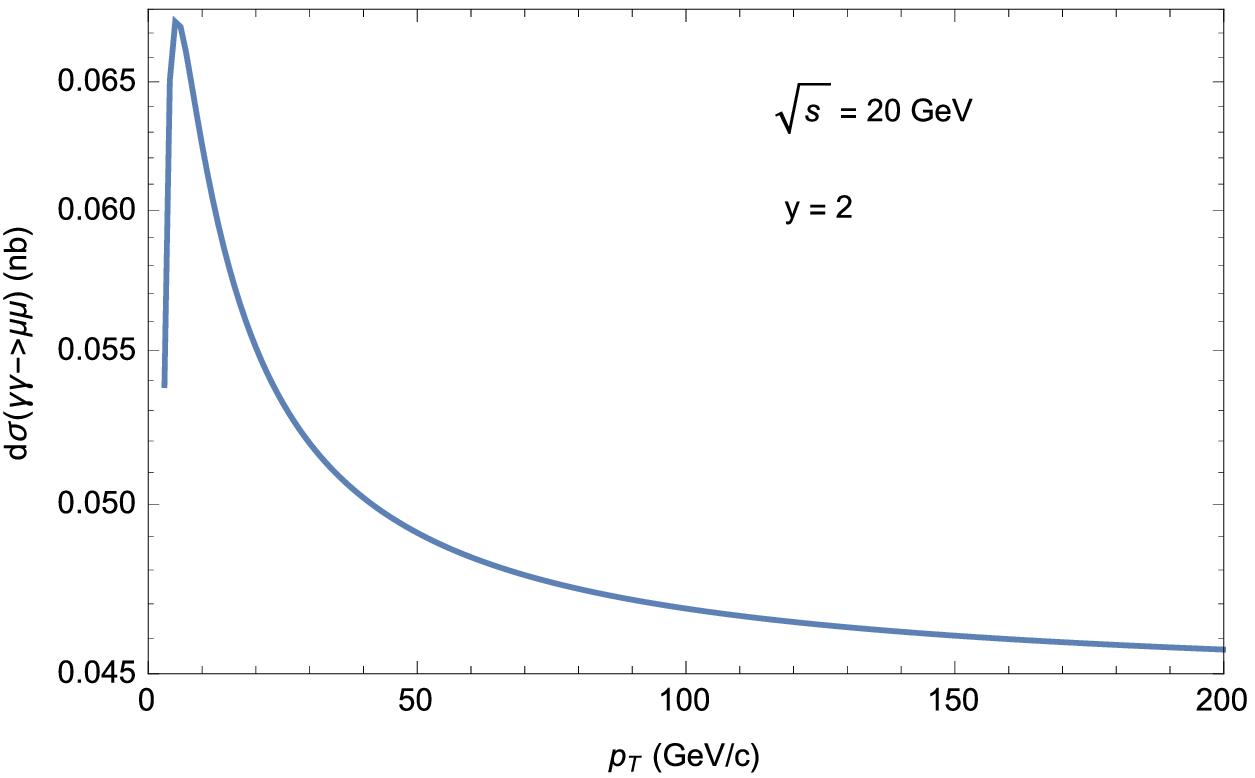}
\parbox{20mm} {\qquad \qquad (c) }
\parbox{9mm} {~~~}
\parbox{40mm} {\hspace*{0.15cm}} {\qquad \qquad (d) }
      \caption{Transverse momentum distribution of $\mu^{\pm}$ for fixed value of centre-of-mass energy $\sqrt {s}$ and for different values of rapidity $y$. ($a$): $\sqrt {s}$ = 20 GeV, $y$ =0; \,\,($b$): $\sqrt {s}$ = 20 GeV, $y$ =1; \,\, ($c$): $\sqrt {s}$ = 20 GeV, $y$ = - 1;
      \,\, ($d$): $\sqrt {s}$ = 20 GeV, $y$ =2.}
       \label{pt2}
\end{figure}

Now we will study the distribution by the transverse momenta and rapidity of muons in the final state in the process \eqref{gamma}.
Therefore, the Mandelstam variables are related to the transverse momentum and rapidity.
Then, the  Mandelstam invariants of the $\gamma\gamma \to \mu^+ \mu^-$ processes satisfy  the following form:
\ba
t &=& m_{\mu}^2 - \sqrt {s}\sqrt {p_{T}^2 + m_{\mu}^2} e^{-y}, \nn \\
u &=& m_{\mu}^2 - \sqrt {s}\sqrt {p_{T}^2 + m_{\mu}^2} e^{y},
\ea
where $p_T$ and $y$ are denote the transverse momentum and rapidity of the muon, respectively. \\
In the center-of-mass system the transverse momentum $\mu^-$ and $\mu^+$ are obviously just opposite
\ba
p_T \equiv p_T^{\mu^-} = -p_T^{\mu^+}.
\ea
The master formula for the differential cross section on the transverse momentum distribution has the form:
\ba
\frac{d\sigma}{dp^2_{T}} = \int dy \cdot \frac{d\sigma}{dp^2_{T} dy}.
\label{PDS}
\ea
The integration limits for the transverse momentum are determined as follows
\ba
0 \leq p_T \leq \frac{1}{2}\sqrt {s -4 m_{\mu}^2},
\ea
from here can see, that $p_T^{\text {max}} = \sqrt {s -4 m_{\mu}^2}$. \\
The differential cross section for the rapidity distribution we can be written in the following form:
\ba
\frac{d\sigma}{dy} = \int dp^2_{T} \cdot \frac{d\sigma}{dy dp^2_{T}}.
\label{YDS}
\ea
The kinematic limits of the rapidity in \eqref{PDS} is determined these relations is given by
\ba
-\ln\frac{2\sqrt {s}}{\sqrt {s-4 m^2_{\mu}}} \leq y \leq \ln\frac{2\sqrt {s}}{\sqrt {s-4 m^2_{\mu}}}.
\ea
In the most common case of equal masses in the final state the differential cross section for the angular distribution has in the following form:
\ba
\frac{d\sigma}{dcos\theta} = \frac{1}{32\pi s}\beta_{\mu} \overline{|{\mathcal M}|^2}.
\label{ADS}
\ea
One of the interesting tasks, in order to, of studying the production of muon pairs in a photon-photon collision is to investigate
the dependence of the cross-section on the muon transverse momentum.
Therefore, we studied in detail the dependence of the differential cross-section on the muon transverse momentum in the
$\gamma\gamma \to \mu^+ \mu^-$ \eqref{gamma} process. \\
In Fig.~\ref{pt1} and Fig.~\ref{pt2} we present our results the investigation of the  muon transverse momentum for different values of centre-of-mass energy and rapidity for the process $\gamma\gamma \to \mu^+ \mu^-$ \eqref{gamma}. \\
In Fig.~\ref{pt1} we the plotted the dependence of the differential cross section for the process $\gamma\gamma \to \mu^+ \mu^-$ \eqref{gamma}
as a function of $p_T$ in the range $2\,\, GeV/c < p_T < 200\,\, GeV/c$ for fixed values values of centre-of-mass energy $\sqrt {s}$ = 10 GeV,
and rapidity $y$ =-1, 0, 1, 2.
In Fig.~\ref{pt1}(a) at a values of rapidity $y$=0, and of $p_{T}$ from 5 Gev/c to 10 GeV/c the differential cross section increases very much,
and at $p_{T}$ = 10 GeV/c the differential cross section reaches a peak value.
After obtaining a peak, with an increase in $p_{T}$ until $p_{T}$ = 200 GeV/c the differential cross section slowly is decreases.
On the other curves in Fig.~\ref{pt1}, that is, in Fig.~\ref{pt1} (b), (c), we plotted the differential cross section as a function of the muon
transverse momentum in the range $3\,\, GeV/c \leq p_T \leq 200\,\, GeV/c$ at rapidity $y$ =1 and $y$ =-1.
The same can be explained, that is, at a value of $p_{T}$ from 3 Gev/c to 7 GeV/c the differential cross section increases very much,
and at $p_{T}$ = 7 GeV/c the differential cross section reaches a peak value, further with  increase in $p_{T}$ until $p_{T}$ = 200 GeV/c,
the cross section slowly is decreases.
Also we plotted in Fig.~\ref{pt1}(d) from the transverse momentum distribution in the range $2\,\, GeV/c \leq p_T \leq 200\,\, GeV/c$
at a values rapidity $y$ =2 and $\sqrt {s}$ = 10 GeV.
It should be noted that at a value of $p_{T}$ from 2 Gev/c to 3 GeV/c the differential cross section the increases sharply,
in the value  $p_{T}$ = 3 GeV the differential cross section reaches a peak value.
In further, in the region $3\,\, GeV/c \leq p_T \leq 30\,\, GeV/c$ the differential cross section decreases is sharply, and then with
increase in $p_{T}$ until $p_{T}$ = 200 GeV/c, the differential cross section slowly is decreases. \\
We present the obtained results in Fig.~\ref{pt2}.
In Fig.~\ref{pt2}(a) at a values of rapidity $y$ =0, and of $p_{T}$ from 8 Gev/c to 20 GeV/c the differential cross section increases very much,
and at $p_{T}$ = 20 GeV/c the differential cross section reaches a peak value.
After obtaining a peak, with an increase $p_{T}$ in region $8\,\,GeV/c \leq p_{T} \leq 200\,\,GeV/c$ the differential cross section slowly is decreases.
On the other curves in Fig.~\ref{pt2}, that is, in Fig.~\ref{pt2} (b), (c), we plotted the differential cross section as a function of the muon transverse momentum in the range $5\,\, GeV/c \leq p_T \leq 200\,\, GeV/c$ at rapidity $y$ =1 and $y$ =-1.
The same can be explained, that is, at a value of $p_{T}$ from 5 Gev/c to 13 GeV/c the differential cross section increases very much,
and at $p_{T}$ = 13 GeV/c the differential cross section reaches a peak value, further with  increase $p_{T}$ in region
$13\,\,GeV/c \leq p_{T} \leq 200\,\,GeV/c$ the cross section slowly is decreases.
Also we plotted in Fig.~\ref{pt2}(d) from the transverse momentum distribution in the range $3\,\, GeV/c \leq p_T \leq 200\,\, GeV/c$ at a values rapidity y=2 and $\sqrt {s}$ = 20 GeV.
It should be noted that at a value of $p_{T}$ from 3 Gev/c to 5 GeV/c the differential cross section the increases sharply, in the value
$p_{T}$ = 5 GeV the differential cross section reaches a peak value.
In farther, in the region $5\,\, GeV/c \leq p_T \leq 40\,\, GeV/c$ the differential cross section decreases is sharply, and then with
increase in $p_{T}$ until $p_{T}$ = 200 GeV/c, the differential cross section slowly is decreases.

\begin{figure}[!htb]
       \includegraphics[width=0.48\linewidth]{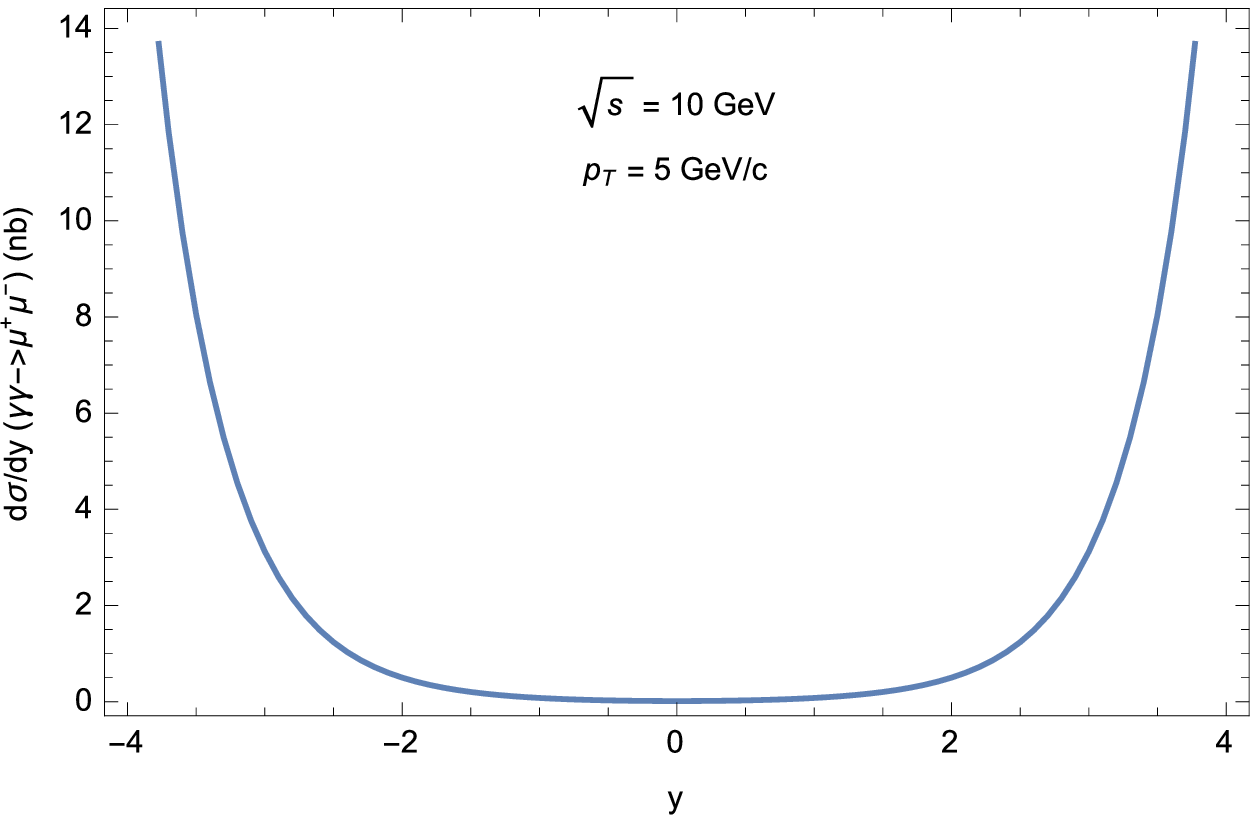}
\hspace*{0.2cm}
       \includegraphics[width=0.48\linewidth]{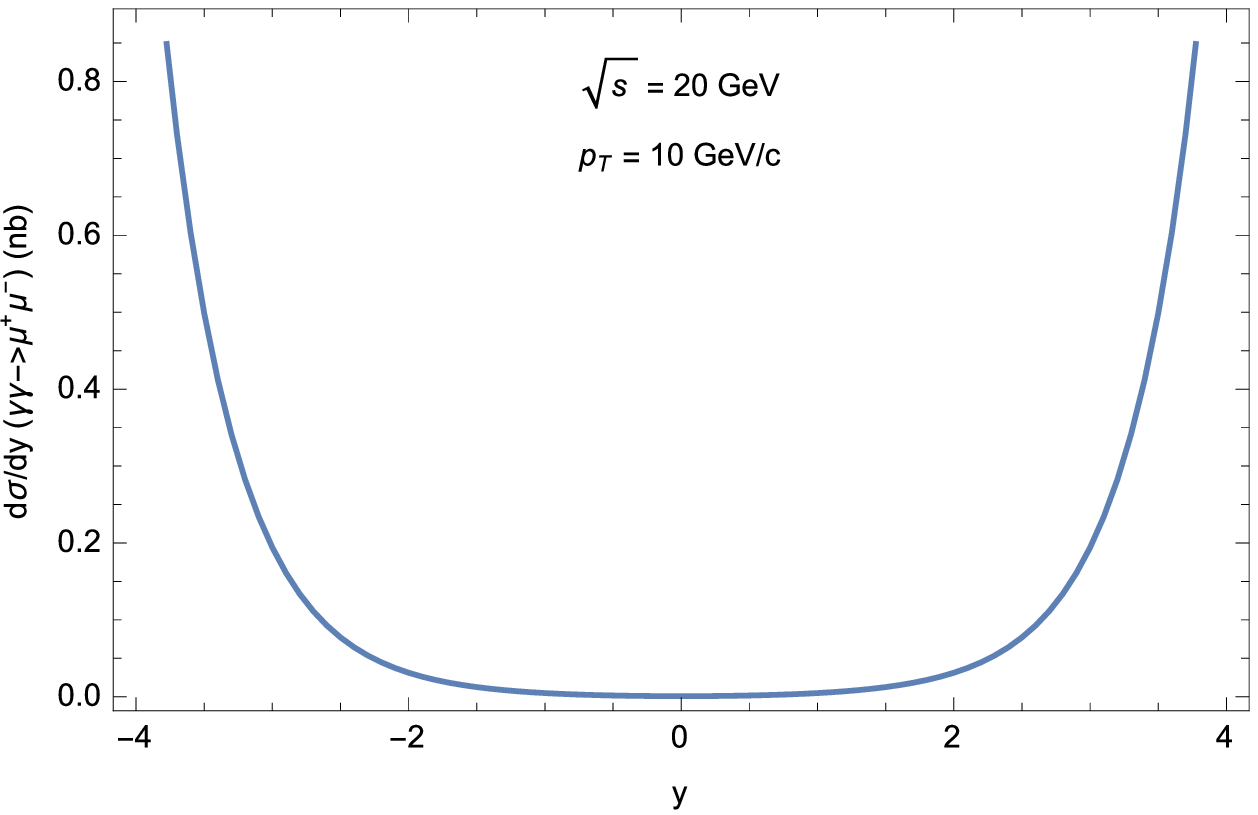}
\parbox{20mm} {\qquad \qquad (a) }
\parbox{9mm} {~~~}
\parbox{40mm} {\hspace*{0.15cm}} {\qquad \qquad (b) }
 \caption{The dependence of differential cross section on rapidity $y$ for different values of centre-of-mass energy $\sqrt {s}$ and
    of transverse momentum $p_{T}$. ($a$): $\sqrt {s}$ = 10 GeV, $p_{T}$ = 5 GeV/c; \,\,($b$): $\sqrt {s}$ = 20 GeV, $p_{T}$ = 10 GeV/c.}
       \label{y}
\end{figure}

Also, we investigate the dependence on process $\gamma\gamma \to \mu^+ \mu^-$ \eqref{gamma} as a function of the rapidity
$y = y_{\mu^+} =y_{\mu^-}$ of the muon pairs in the range $-4 \leq y \leq 4$ for different values of the center-of-mass energy
$\sqrt {s}$ and of the transverse momentum $p_T$. \\
The rapidity distributions we demonstrates in the Fig.~\ref{y}.
In Fig.~\ref{y}(a), we have depicted differential cross section as a function of the muon rapidity $y$ at $\sqrt {s} = 10\,\, GeV$ and
$p_{T} = 5\,\, GeV/c$. In the region ($-4 \leq y \leq -2$) the differential cross section decreases sharply, but in the region
($-2 \leq y \leq 2$) the differential cross section changes very slowly (the practically stable).
It is seen that  we have a minimum value of the differential cross section at the point $y = 0$.
In farther, in the region ($2 \leq y \leq 4$) the differential cross section the increases sharply. \\
The analogous calculations we have done for $\sqrt {s}$ = 20 GeV and $p_{T}$ = 10 GeV/c.
Obtained our results is present in Fig.~\ref{y}(b).
It is note that for these values of $\sqrt {s}$ = 20 GeV and $p_{T}$ = 10 GeV/c, the same, in the region ($-4 \leq y \leq -2$) the
differential cross section decreases sharply, but in the region ($-2 \leq y \leq 2$) the differential cross section changes very slowly
(the practically stable). It is worth to mention that at the $y=0$ the differential cross section the obtain a minimum value.
In farther, in the region ($2 \leq y \leq 4$) the differential cross section the increases sharply.

\begin{figure}[!htb]
       \includegraphics[width=0.48\linewidth]{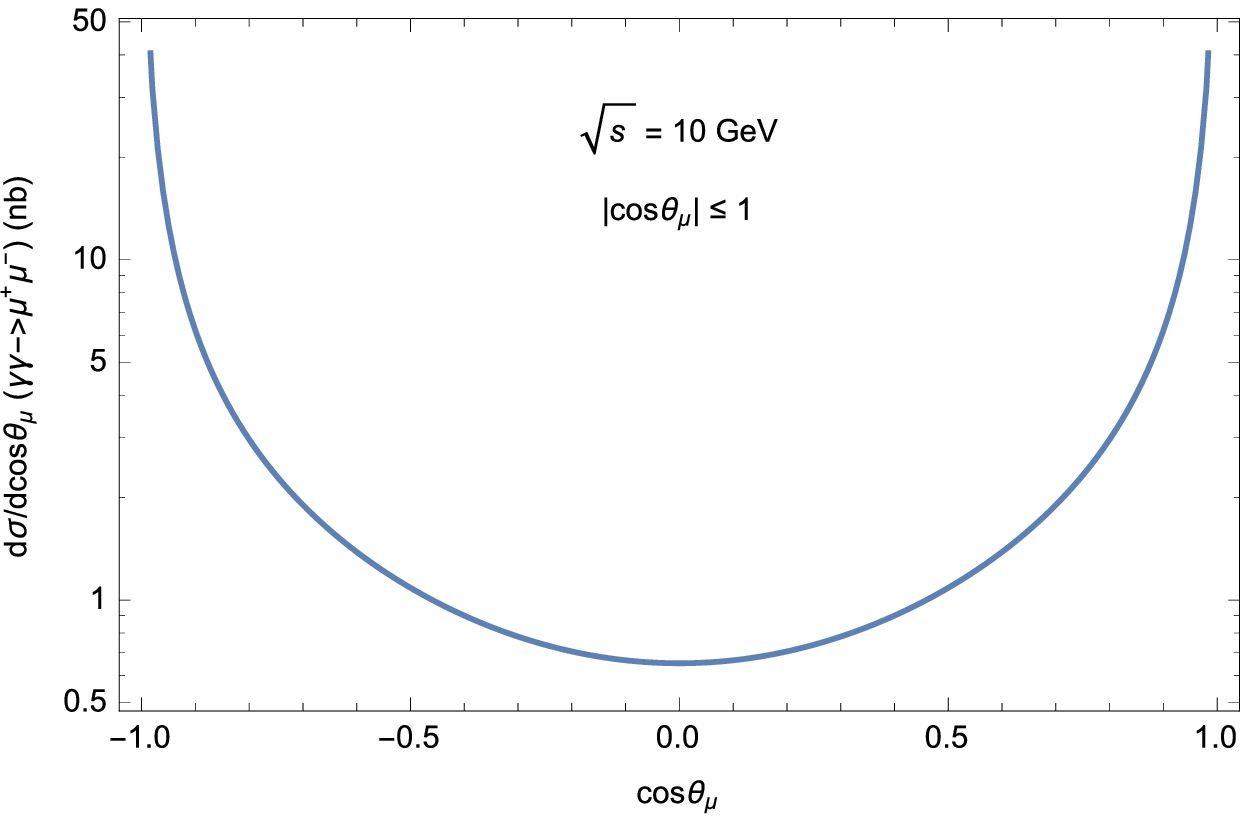}
  \hspace*{0.05cm}     \includegraphics[width=0.48\linewidth]{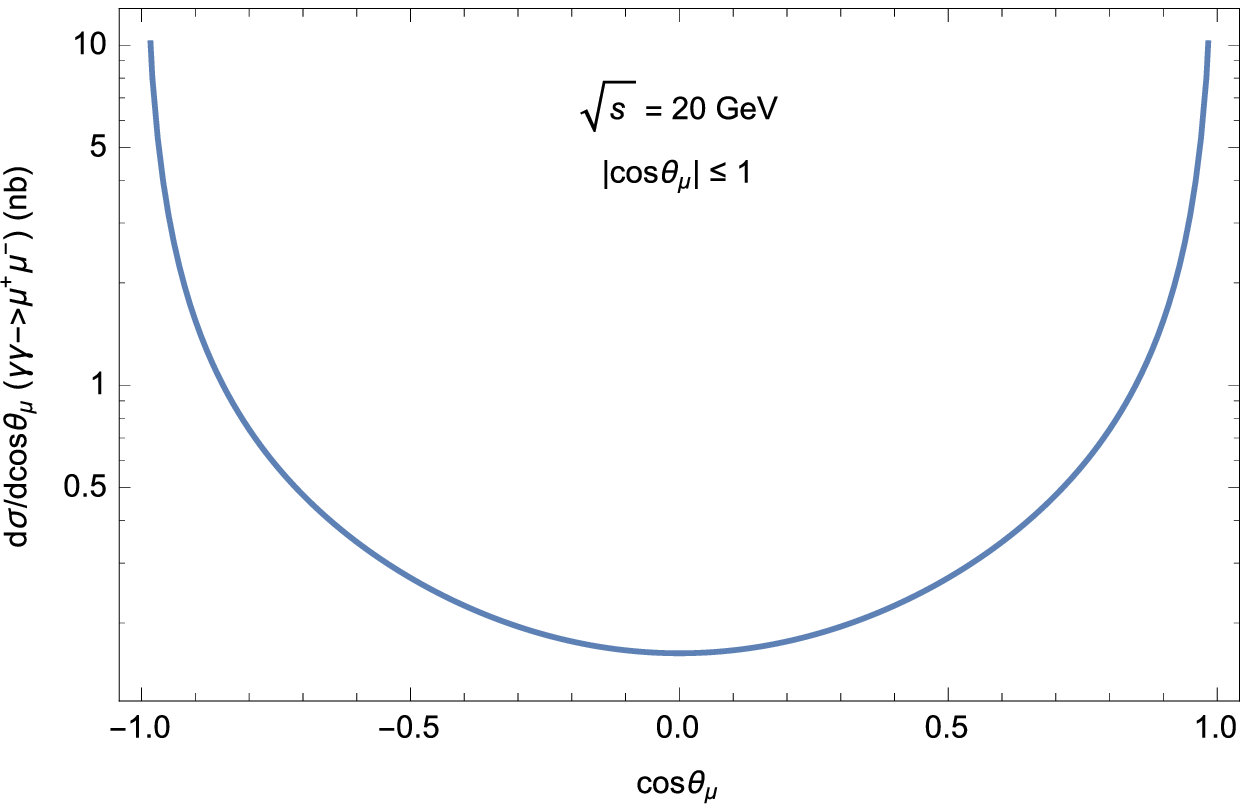} \\
\parbox{20mm} {\qquad \qquad (a) }
\parbox{9mm} {~~~}
\parbox{40mm} {\hspace*{0.15cm}} {\qquad \qquad (b) }

\vspace*{0.5cm}
\centering
 \hspace*{2.0cm}
    \includegraphics[width=0.48\linewidth]{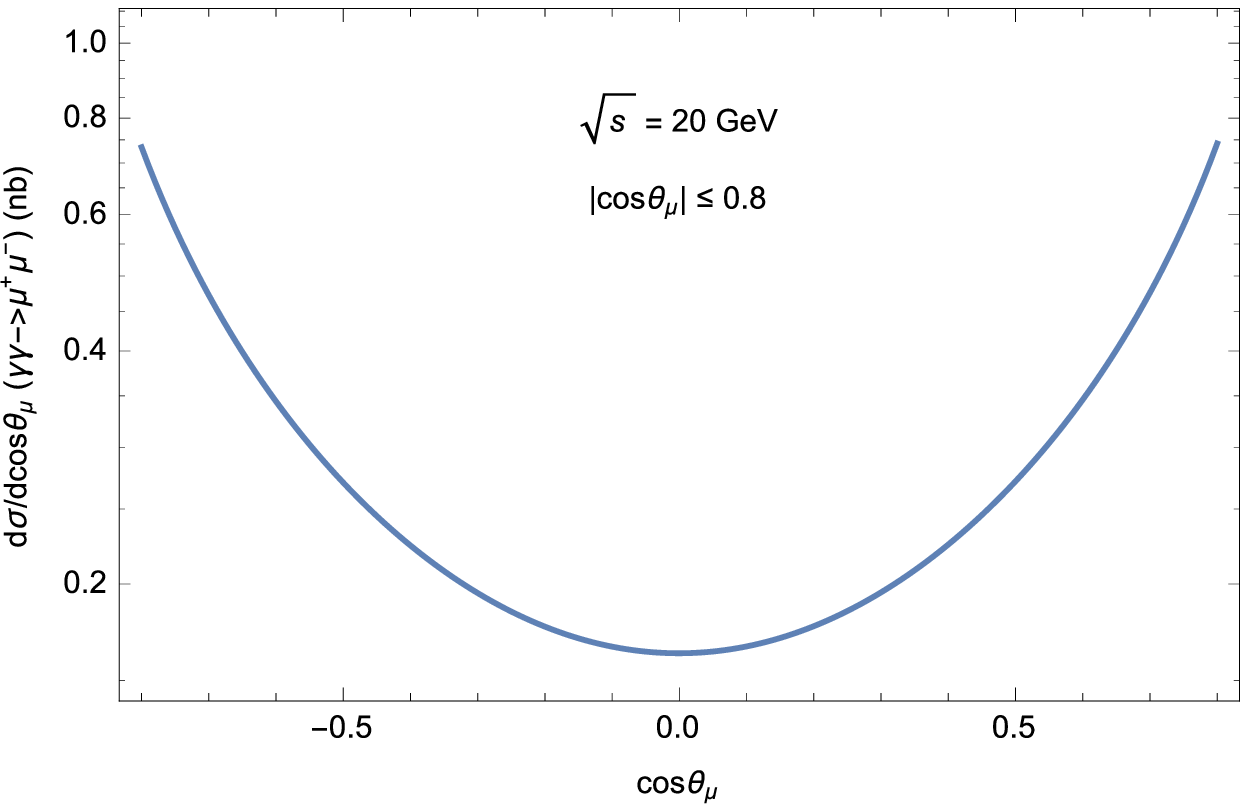}

\vspace*{0.01cm}
\parbox{15mm} {\hspace*{0.15cm}} {\qquad \qquad (c) }
       \caption{The angular distribution of the differential cross section of the process $\gamma\gamma \to \mu^+ \mu^-$
       for different values of centre-of-mass energy $\sqrt {s}$ and of angle $cos\theta_{\mu}$.}
       \label{cos}
\end{figure}

One of the interesting problem, of the muon pairs production in a photon-photon collisions is investigate the dependence of the
angular distribution of the differential cross-section. Therefore, we studied the dependence of the differential cross-section on the
muon angular distribution by the \eqref{ADS} in the $\gamma\gamma \to \mu^+ \mu^-$ \eqref{gamma} process.
In Fig.~\ref{cos} we show the dependence of the differential cross section for the process $\gamma\gamma \to \mu^+ \mu^-$ \eqref{gamma}
as a function of cos $\theta_{\mu}$ in the ranges (-1 $\leq$ cos $\theta_{\mu} \leq$ 1) and (-0.8 $\leq$ cos $\theta_{\mu} \leq$ 0.8)
for for different values of centre-of-mass energy $\sqrt {s}$.
In Fig.~\ref{cos}(a), we have plotted differential cross section as a function of the muon angular distribution in the region
(-1 $\leq$ cos $\theta_{\mu} \leq$ 1) at $\sqrt {s} = 10\,\, GeV$.
In the region (-1 $\leq$ cos $\theta_{\mu} \leq$ -0.2) the differential cross section decreases sharply, but in the region
(-0.2 $\leq $ cos $\theta_{\mu} \leq$ 0.2) the differential cross section changes very slowly.
At the value $\theta_{\mu} = 90^{\circ}$ the differential cross section obtain a minimum value.
It is seen that in further in the region (0.2 $\leq $ cos $\theta_{\mu} \leq$ 1) the differential cross section the increases sharply. \\
The analogous calculations we have done for $\sqrt {s}$ = 20 GeV, and this dependence is displayed in Fig.~\ref{cos}(b).
It is shown that in the region (-1 $\leq$ cos $\theta_{\mu} \leq$ -0.2) the differential cross section decreases sharply, but in the region
(-0.2 $\leq $ cos $\theta_{\mu} \leq$ 0.2) the differential cross section changes very slowly.
At the value $\theta_{\mu} = 90^{\circ}$ the differential cross section obtain a minimum value.
The same that in further in the region (0.2 $\leq $ cos $\theta_{\mu} \leq$ 1) the differential cross section the increases sharply. \\
It should be noted that in the experiment, the angular distribution of the differential cross section also measurement in the region
$|cos \theta_{\mu}| \leq 0.8$.
Therefore, the analogous calculations we have done for the angular distribution of the differential cross section in the region
(-0.8 $\leq $ cos $\theta_{\mu} \leq$ 0.8) at $\sqrt {s}$ = 20 GeV.
Obtained results of our numerical calculation is plotted in Fig.~\ref{cos}(c).
It is shown that in the region (-0.8 $\leq$ cos $\theta_{\mu} \leq$ -0.1) the differential cross section decreases sharply, but in the region
(-0.1 $\leq $ cos $\theta_{\mu} \leq$ 0.1) the differential cross section changes very slowly.
At the value $\theta_{\mu} = 90^{\circ}$ the differential cross section obtain a minimum value.
In farther, in the region (0.1 $\leq $ cos $\theta_{\mu} \leq$ 0.8) the differential cross section the increases sharply.

%
%
%
%

\vspace*{0.048cm}
\section{The process $\bf {e^+ + e^- \to e^+ + e^- + \mu^+ + \mu^-}$}\label{ir}

When an electron and a positron interact with each other, many different phenomena occur.
One of these phenomena is process of the creation of $\mu^+\mu^-$ pairs. \\
One of the most important areas at the QED and SM are investigation of the production massive
lepton pairs at the electron-positron linear collider.

In this paper, we report results from a study of $\alpha^4$ order
QED process $e^+e^- \to e^+e^-\mu^+\mu^-$ - at center-of mass
energies from $161\,\, GeV$ to $209\,\, GeV$. \\
The lepton pairs production in electron-positron collisions by a two photon mechanism, play a special role in QED and QCD \cite{Budnev},
since their analysis is under much better control. \\
Therefore, it should be noted that the production of lepton pairs in electron-positron collisions via photon-photon-fusion has been studied for a long time.

\begin{figure}[!htb]
\includegraphics[width=0.43\linewidth]{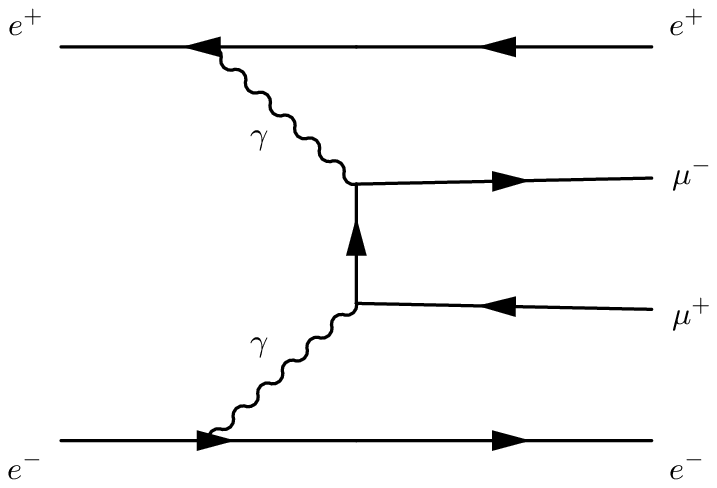}
\hspace*{1cm}
\includegraphics[width=0.43\linewidth]{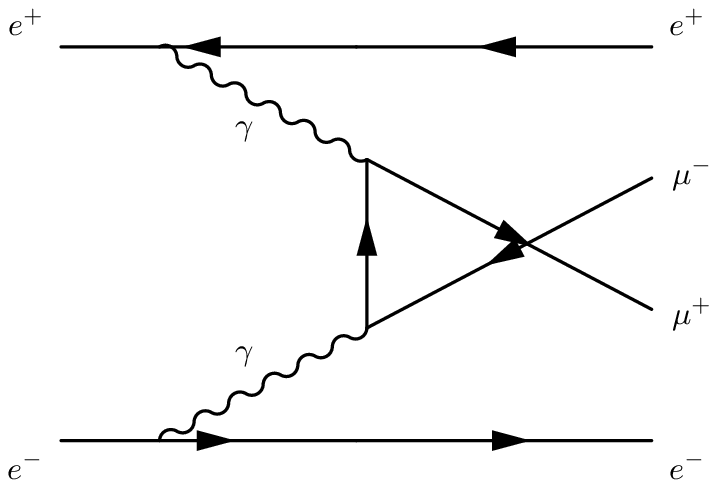}
\parbox{20mm} { (1) }
\parbox{30mm} {~~~ }
\parbox{50mm} { \qquad \qquad  \qquad  (2) }
       \caption{Representative Feynman diagrams for the process of $e^+ + e^- \to e^+ + e^- + \mu^+ + \mu^-$.}
       \label{diagram8}
\end{figure}

The process of the production of lepton pairs in electron-positron collisions  via photon-photon-fusion can be written in the form
\ba
e^+ + e^- \to e^+ + \gamma^{\ast} + e^- + \gamma^{\ast} \to e^+ + e^- + \mu^+ + \mu^-.
\label{eemm}
\ea
The Feynman diagrams of the process \eqref{eemm} can be represented in Fig.~\ref{diagram8}.

The total cross section of the $e^+e^- \to e^+e^-\mu^+\mu^-$ process, which we study through photon-photon-fusion sub-process can be written as
\cite{Budnev,gamma1}
\ba
\sigma(s)^{e^+e^- \to e^+e^- \mu^+ \mu^-} = \biggl(\frac{\alpha}{\pi}\biggr)^2
\int\limits_{4m_{\mu}^2}^{4E^2}\frac{ds_1}{s_1}\sigma^{\gamma\gamma \to \mu^+ \mu^-}(s_1)\biggl[\biggl(\ln\frac{sm_{\mu}^2}{s_1 m_e^2} - 1\biggr)^2 f\biggl(\frac{s}{s_1}\biggr)-
\frac{1}{3}\biggl(\ln\frac{s}{s_1}\biggr)^3\biggr],\,\,\,\,\,\,\,
\label{cseemm}
\ea
where
\ba
f(x) = \biggl(1+\frac{1}{2x}\biggr)^2\ln x - \frac{1}{2}\biggl(1-\frac{1}{x}\biggr)\biggl(3+\frac{1}{x}\biggr),
\label{f}
\ea
where $\sigma^{\gamma\gamma \to \mu^+ \mu^-}(s_1)$ is the cross section of the $\gamma\gamma \to \mu^+ \mu^-$ process.
\begin{figure}[!htb]
       \centering
       \includegraphics[width=0.48\linewidth]{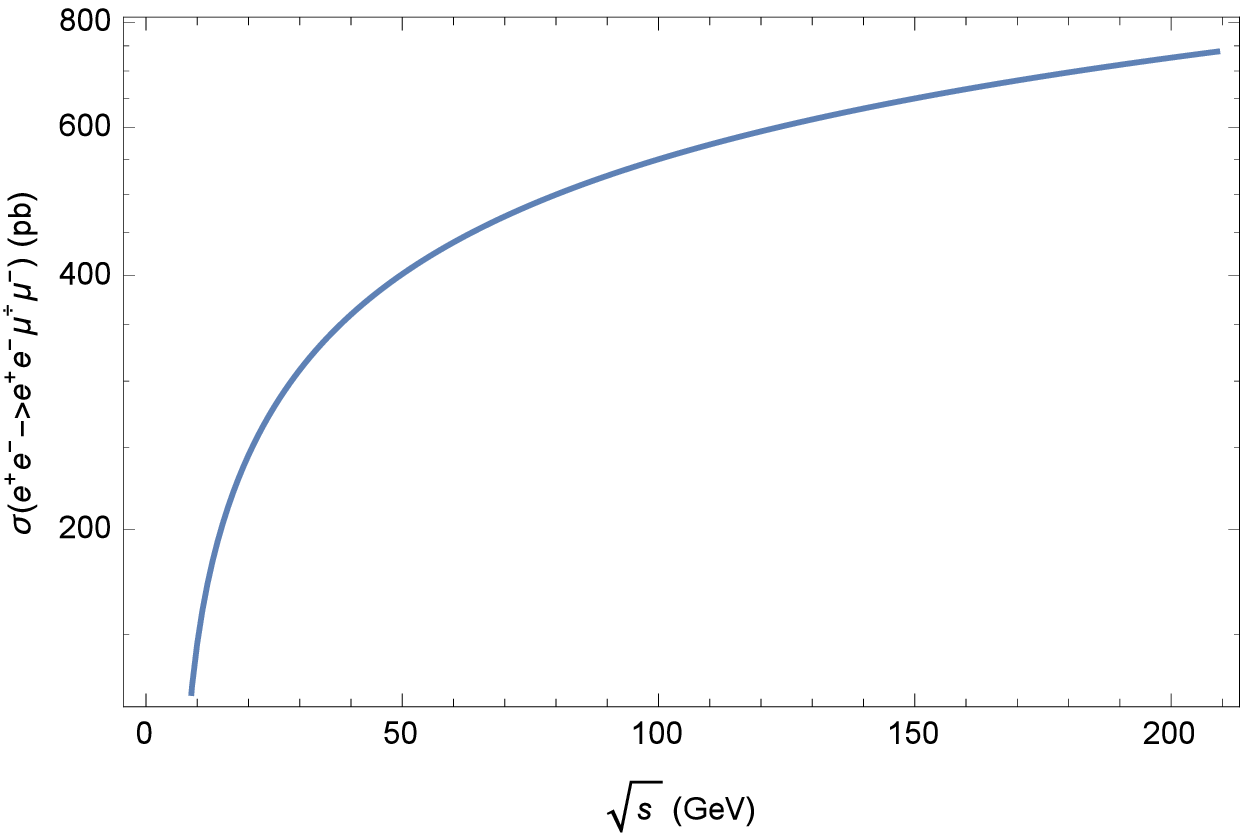}
       \includegraphics[width=0.48\linewidth]{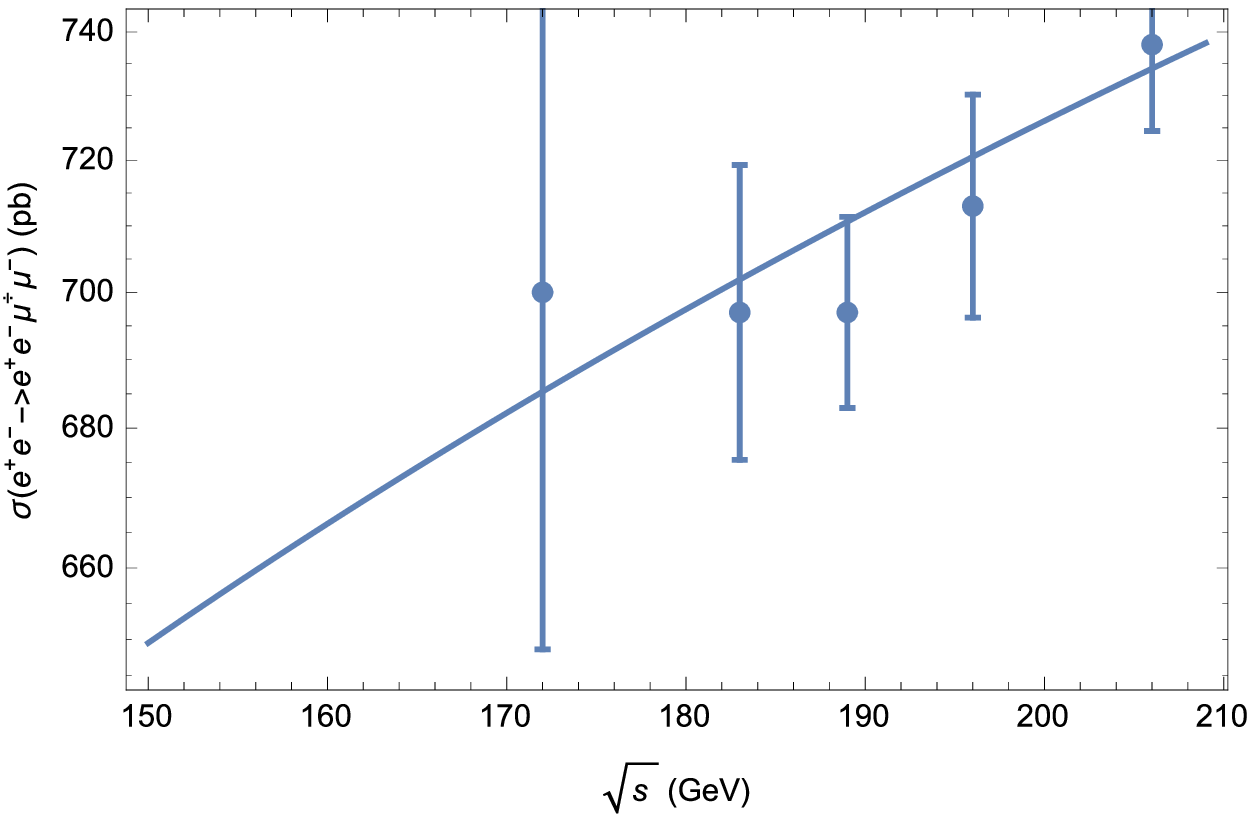} \\
\parbox{20mm} {\qquad \qquad (a) }
\parbox{9mm} {~~~}
\parbox{40mm} {\hspace*{0.15cm}} {\qquad \qquad (b) }

\vspace*{0.5cm}
    \hspace*{0.15cm}   \includegraphics[width=0.48\linewidth]{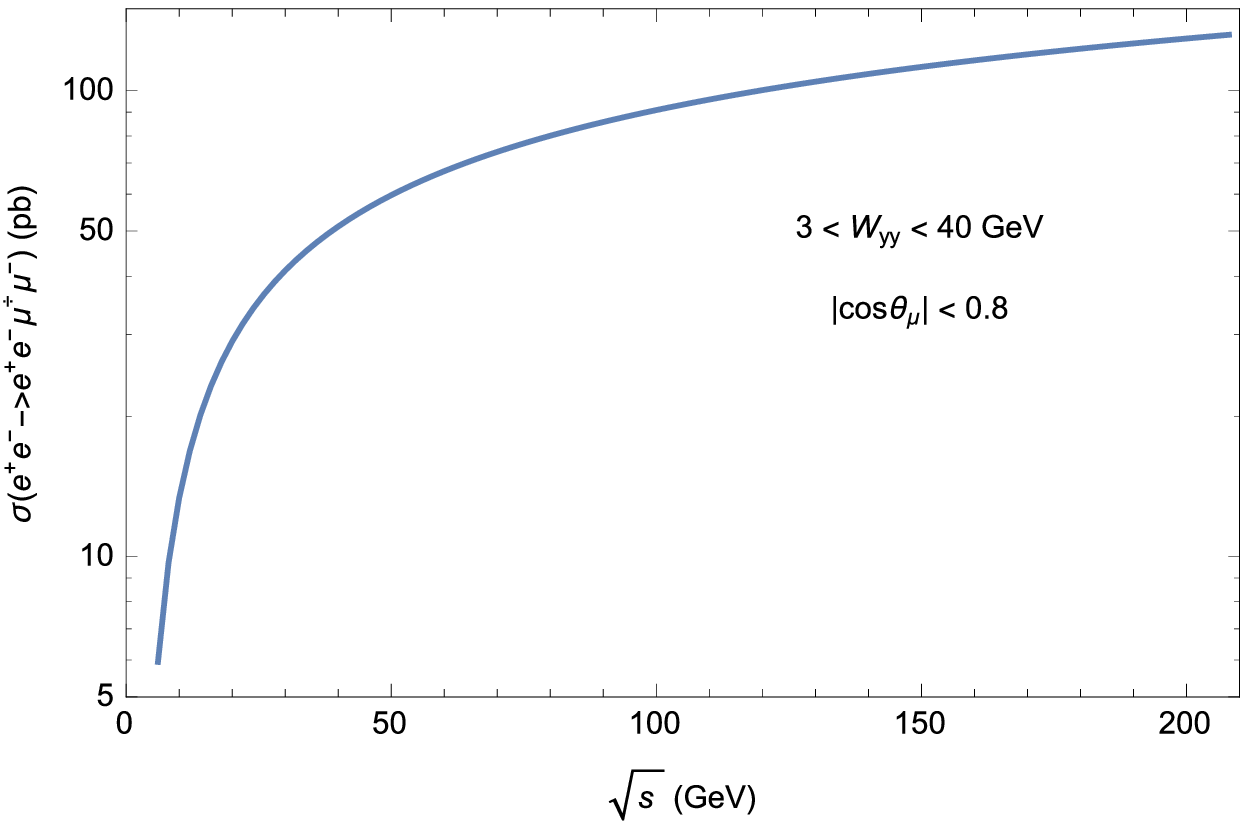}
     \hspace*{0.05cm}  \includegraphics[width=0.48\linewidth]{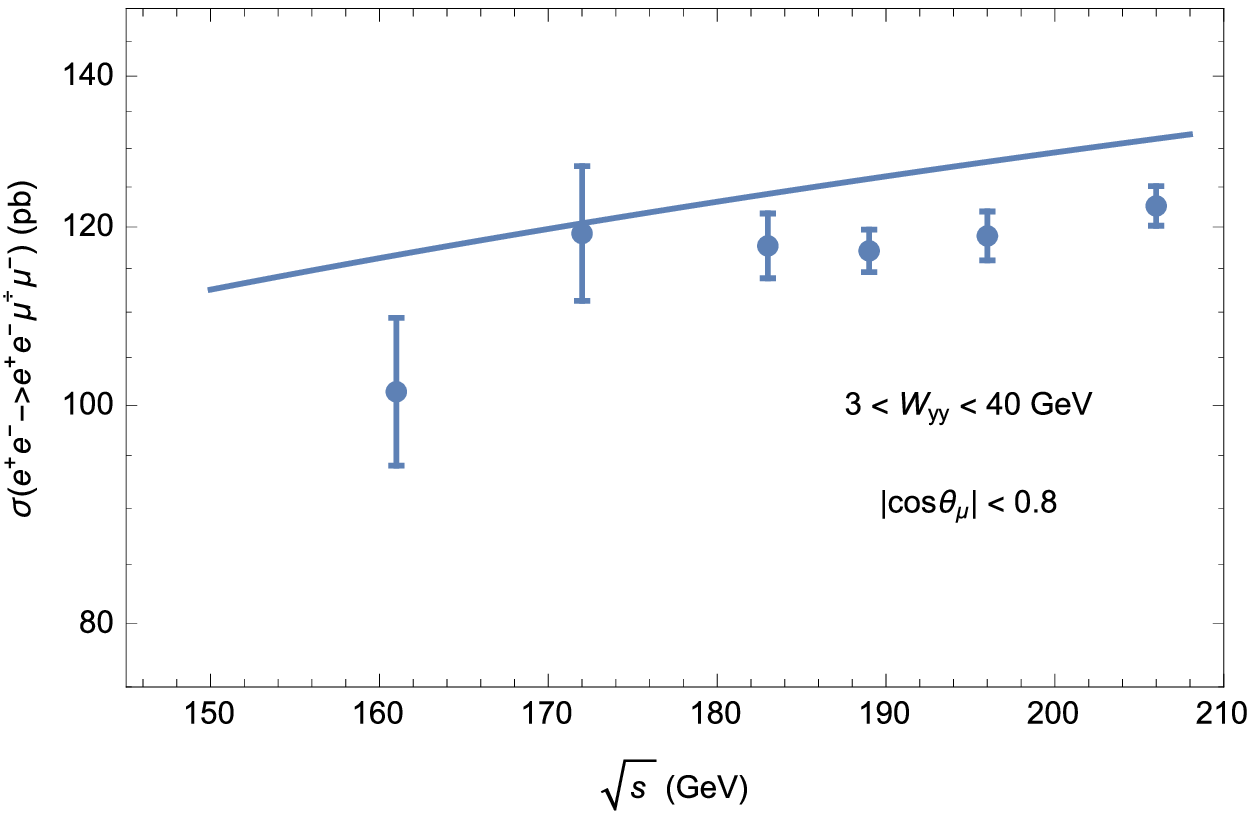}
\parbox{20mm} {\qquad \qquad (c) }
\parbox{9mm} {~~~}
\parbox{40mm} {\hspace*{0.15cm}} {\qquad \qquad (d) }
      \caption{The total cross section of the $e^+ + e^- \to e^+ + e^- + \mu^+ + \mu^-$ process as a function of $\sqrt {s}$.
       The experimental data is from \cite{L3} of the L3 Collaboration.
       ($a$): $6\,\, GeV\leq \sqrt {s} \leq 209\,\, GeV$; \,\,($b$): $100\,\, GeV\leq \sqrt {s} \leq 209\,\, GeV$ - for the compared with experimental data; \,\, ($c$): $3\,\, GeV\leq W_{\gamma\gamma} \leq 40\,\, GeV$ and $|cos\theta_{\mu}| \leq 0.8$, \,\,;
      \,\, ($d$): $150\,\, GeV\leq \sqrt {s} \leq 209\,\, GeV$,\,\, $3\,\, GeV\leq W_{\gamma\gamma} \leq 40\,\, GeV$ and $|cos\theta_{\mu}| \leq 0.8$ \,\,- for the compared with experimental data.}
       \label{TCSee}
\end{figure}

Using the two-photon mechanism, we calculated the total cross section of the $e^+ + e^- \to e^+ + e^- + \mu^+ + \mu^-$ process \eqref{eemm}
according to formula \eqref{cseemm} in a large range of centre-of-mass energy $\sqrt {s}$ and for energy that corresponds to an experiment
with a specific angle range. In Fig.~\ref{TCSee}, the dependence of the total cross section on the center-of-mass energy of the
$e^+ + e^- \to e^+ + e^- + \mu^+ + \mu^-$ process are plotted.
We present our results the dependence the total cross section, as a function of the $e^+ e^-$ center-of-mass energy
$\sqrt {s}$ in the region 5 GeV $\leq \sqrt {s} \leq$ 209 GeV in Fig.~\ref{TCSee}(a).
According to our calculation, we see that total cross sections in the region of the center-of-mass energy 5 GeV $\leq \sqrt {s} \leq$ 150 GeV
increase sharply.
In farther, total cross sections increase smoothly and slowly with increasing energy in the region 150 GeV $\leq \sqrt {s} \leq$ 209 GeV.
Accordingly with experimental data \cite{L3}, we made a analogous calculation for a specific range of the center-of-mass energy
160 GeV $\leq \sqrt {s} \leq$ 209 GeV, in order to compare with this experimental data.
In Fig.~\ref{TCSee}(b) we show the our obtained results.
In Fig.~\ref{TCSee}(b) it is seen that our theoretical results are in satisfactory agreement with the experimental data. \\
We also investigated the dependence of the cross section as a function of $\sqrt {s}$ in the $e^+ + e^- \to  e^+ + e^- + \mu^+ + \mu^-$
\eqref{eemm} process for the case two-photon center-of-mass energy $3\,\, GeV \leq W_{\gamma\gamma} \leq 40\,\, GeV$
and of muon angles in the range $|cos \theta_{\mu}| \leq 0.8$.
Therefore, in formula \eqref{cseemm} for the $\gamma\gamma \to \mu^+ \mu^-$ subprocess we use formula \eqref{ADS} and we are perform of
integration on the $W_{\gamma\gamma}$ in the range $3\,\, GeV \leq W_{\gamma\gamma} \leq 40\,\, GeV$ and on the muon angle in the region
(-0.8 $\leq $ cos $\theta_{\mu} \leq$ 0.8). In Figs.~\ref{TCSee}(c) and ~\ref{TCSee}(d), we present our obtained results for these cases.
We plot the dependence of the cross section of $e^+ + e^- \to  e^+ + e^- + \mu^+ + \mu^-$
\eqref{eemm} process as a function for $e^+ e^-$ centre-of-mass energy in the region 5 GeV $\leq \sqrt {s} \leq$ 209 GeV at the
$3\,\, GeV \leq W_{\gamma\gamma} \leq 40\,\, GeV$ and (-0.8 $\leq $ cos $\theta_{\mu} \leq$ 0.8) in Fig.~\ref{TCSee} (c).
Of our calculation it is shown that total cross sections in the region of the $e^+ e^-$ center-of-mass energy
5 GeV $\leq \sqrt {s} \leq$ 110 GeV increase sharply. In farther, total cross sections increase smoothly and slowly with increasing
energy in the region 110 GeV $\leq \sqrt {s} \leq$ 209 GeV.
For the compare with experimental data \cite{L3}, we are perform of analogous calculation for a specific range of the center-of-mass energy
160 GeV $\leq \sqrt {s} \leq$ 209 GeV.
In Fig.~\ref{TCSee}(d) it is shown that our theoretical results are in satisfactory agreement with the experimental data.

\section{Conclusion}
\label{Conclusion}

Are known, to be the photon-photon and the electron-positron collisions the most elementary interactions and form the basis of our knowledge about
the nature of high energy physics. \\
The exclusive production of $\mu^+ \mu^-$ pair in electron-positron collisions via two-photon mechanism is considered.
In this process, we present a simple method for the calculation of the total and differential cross section.

A careful study of the muon pairs production reactions over the full range of LEP, LHC, BABAR, BELLE, Fermilab et al. $\sqrt {s}$
values and high $p_T$ is highly desirable, both as an important probe, as a significant test in SM. \\
After a detailed study of the photon-photon-collision processes, one can notice that two-photon reactions have of unique features for
testing QED and QCD.

For the calculation the dependence total and differential cross section  on the centre-of-mass energy, the  ransverse momentum,
and the rapidity and the angle, we got the master formula for the $\gamma\gamma \to \mu^+\mu^-$ process.

It should be noted that the total cross section of the $e^+e^- \to e^+e^- \mu^+\mu^-$ process was measured of the two-photon
center-of-mass energy in the region of 3 GeV $\leq W_{\gamma\gamma} \leq$ 40 GeV by using the data, in which taken from the L3
detector at 161 GeV $\leq \sqrt {s} \leq$ 209 GeV at LEP.  \\
We also want to note that it should be noted that the total cross section muon pairs production as a function of center-of-mass energy
of the $\gamma\gamma \to\mu^+\mu^-$ process was measured of the two-photon center-of-mass energy $\sqrt {s}$ in the range
3 GeV $\leq W_{\gamma\gamma} \leq$ 40 GeV by using the data, from the L3 detector at LEP.

In this present paper, we have examined the muon pairs production in photon - photon collisions and in the process
$e^+e^- \to e^+e^- \mu^+ \mu^-$  by the two-photon mechanism $\gamma\gamma \to \mu^+\mu^-$.\\
We have investigated in detail the dependence total cross section as a function of the two-photon centre-of-mass energy for
3 GeV $\leq W_{\gamma\gamma} \leq$ 40 GeV, and the dependence differential cross section on the transverse momentum $p_{T}$ in the region
2 GeV/c $\leq p_T \leq$ 200 GeV/c for different values of centre-of-mass energy
$\sqrt {s}$=10, 20 GeV, and of rapidity $y$ = -1, 0, 1, 2, in the process of $\gamma\gamma \to \mu^+\mu^-$. \\
We also investigated the dependence of the differential cross section on rapidity $y$ for different values of centre-of-mass energy
$\sqrt {s}$=10, 20 GeV, and of transverse momentum $p_{T}$=5, 10 GeV/c in the process of $\gamma\gamma \to \mu^+\mu^-$.

In this present work, we also was studied in detail the angular distribution of the differential cross section
$\frac{d\sigma (\gamma\gamma \to \mu^+\mu^-)}{dcos\theta_{\mu}}$ of the $\gamma\gamma \to \mu^+\mu^-$ process.
For this, we have separately considered the angle regions in the range of  (-1$ \leq cos\theta_{\mu} \leq$1) and
(-0.8$\leq cos\theta_{\mu} \leq$0.8), and for different values of centre-of-mass energy of the two-photon system $\sqrt {s}$ =10, 20 GeV.

We also investigated the main process for the $\mu^+ \mu^-$ production in the $e^+ + e^- \to e^+ + e^- + \mu^+ + \mu^-$ process and studied the
characteristics of the total cross section, as a function of the $e^+ e^-$ center-of-mass energy $\sqrt {s}$ in the region
5 GeV $\leq \sqrt {s} \leq$ 209 GeV of this process by the two-photon mechanism for different values of angle regions
$-1 \leq cos\theta_{\mu} \leq 1$ and $-0.8 \leq cos\theta_{\mu} \leq 0.8$. \\
We want to note that some of our obtained theoretical results are compare with the experimental data in the processes of
$\gamma\gamma \to \mu^+\mu^-$ and $e^+ + e^- \to e^+ + e^- + \mu^+ + \mu^-$. They are in satisfactory agreement. \\
To make the evaluation for the loop integrals in Box diagrams, we used packages LoopTools \cite{LT} in based on Passarino-Veltman
reduction techniques \cite{PV}. \\
In our opinion of our obtained theoretical results will be helpful for investigations and analysis of the lepton pairs production
in the $e^+ e^-$ and the other colliders.


\end{document}